\documentclass[aps,amsmath,preprint,showpacs]{revtex4}
\usepackage{graphicx}
\usepackage{dcolumn}
\usepackage{color}
\usepackage{epstopdf}
\newcommand*{\s}[1]{/\llap{$#1$}} 
\long\def\Omit#1{}

\renewcommand*{\eqref}[1]{Eq.~(\ref{eq:#1})}
\newcommand*{\eqlab}[1]{\label{eq:#1}}
\newcommand*{\figref}[1]{Fig.~\ref{fig:#1}}
\newcommand*{\figlab}[1]{\label{fig:#1}}
\newcommand*{\secref}[1]{Section~\ref{sec:#1}}
\newcommand*{\seclab}[1]{\label{sec:#1}}

\newcommand{\beq}{\begin{eqnarray}}
\newcommand{\eeq}{\end{eqnarray}}

\newcommand{\Slash}[1]{\ooalign{\hfil/\hfil\crcr$#1$}}

\def\gsim{\displaystyle\mathop{>}_{\sim}}
\def\lsim{\displaystyle\mathop{<}_{\sim}}

\begin{document}

\title{A coupled-channel analysis for $\phi$-photoproduction with $\Lambda$(1520)}
\author{S.\ Ozaki}
\email{sho@rcnp.osaka-u.ac.jp}
\author{A.\ Hosaka}
\email{hosaka@rcnp.osaka-u.ac.jp}
\affiliation{Research Center for Nuclear Physics (RCNP),
Osaka University, Ibaraki, Osaka, 567-0047, Japan}
\author{H.\ Nagahiro}
\email{nagahiro@rcnp.osaka-u.ac.jp}
\affiliation{Department of Physics, Nara Women's University,
Nara 630-8506, Japan}
\author{O.\ Scholten}  \email{scholten@kvi.nl}
\affiliation{Kernfysisch Versneller Instituut, University of Groningen,
9747 AA Groningen, The~Netherlands}
\date{\today}

\begin{abstract}

We investigate photoproduction of $\phi$-mesons off protons within a
coupled-channel effective-Lagrangian method which is based on the K-matrix
approach. Since the threshold energy of the $K\Lambda(1520)$ channel is close to
that of $\phi N$, the contribution of this channel to $\phi$-photoproduction near
the threshold energy region may give rise to some unexpected structures. In the
transition amplitude $K\Lambda(1520) \to \phi N$, the kinematics allows an
intermediate kaon to be on-shell. This happens in the energy region where a peak
structure has been observed in $\phi$-photoproduction. In our calculations the
on-shell kaon effect indeed reproduces a peak structure, however, with a
magnitude that is far too small to explain the observed effect. As a following step
we introduce a nucleon resonance in our model. The coupling of the resonance to
the $K\Lambda(1520)$ and $\phi N$ channels is not suppressed by the OZI rule if
the resonance contains a dominant hidden strangeness component. We find that the
resonance can reproduce a peak structure of the correct magnitude at the right
energy. We also investigate the effects of coupled channels and the resonance on
the angular distribution and the spin density matrices for $\phi$
photoproduction.

\end{abstract}
\maketitle

\newpage

\section{Introduction}

Photo-induced strangeness production is one of main topics in hadron physics. At the relevant
photon energy of about 1 GeV one is still well below the regime of perturbative
QCD and hence one expects large non-pertubative effects. At this energy one is
also well above the energy region that is controlled by low-energy theorems. At
the energies for strangeness production important ingredients are the various baryon resonances and coupled-channel effects.

\begin{figure*}
\begin{center}
\includegraphics[width=0.5 \textwidth]{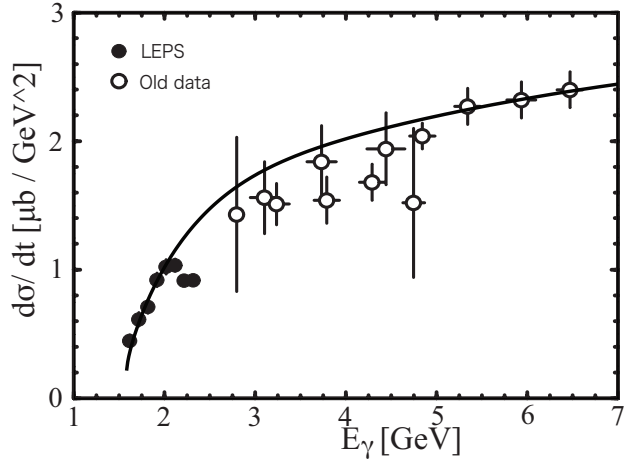}
\vskip -0.1in \caption{The differential cross section $d\sigma/dt \ (\theta = 0)$ as a function of the photon energy $E_{\gamma}$ in the laboratory system. The solid line
is the sum of pomeron and meson($\pi$ and $\eta$) exchange contributions.
The data are taken from Refs.~\cite{Mibe:2005er,Durham}
\label{data}}
\end{center}
\end{figure*}

Recently, several photon facilities have reported
interesting results in the energy region
of strangeness production, such as
pentaquarks~\cite{Nakano:2003qx,Nakano:2008ee},
$\Lambda$ resonances~\cite{Barrow:2001ds,Niiyama:2008rt},
and $\phi$-meson production~\cite{Anciant:2000az,Mibe:2005er,Santoro:2008ai}.
The latter has the unique feature that
the gluon dynamics dominates in the reaction process because
the process is OZI suppressed due to the dominant
$ \bar ss$ structure of the $\phi$-meson.
As shown in Fig.~\ref{data},
the cross section of $\phi$-photoproduction increases with increasing energy which
can be explained by a Pomeron and meson exchange model~\cite{Titov:2003bk,Titov:2007fc}.
The Pomeron is introduced in Regge theory for high-energy hadron scattering
and is considered to be dominated by gluon dynamics.
It was also shown that this model reproduces the
angular dependence in the diffractive region, spin observables and the energy
dependence~\cite{Titov:1999eu}.
Since Regge theory was developed to describe average properties of high energy
scattering processes, such a good agreement is much better than what could have
been expected.
Extrapolation from the high-energy region predicts a smooth energy dependence of the cross section down to the
threshold energy for the reaction.
Interestingly, the LEPS recent observation has shown a strong indication of a peak
structure (blobs in Fig.~\ref{data})
at around $E_\gamma({\rm{Lab}}) \sim 2$~GeV.
Such a structure is difficult to explain
in conventional models of $\phi$-photoproduction, and it is the subject of the
present paper to explore possible explanations for this.

The energy where the peak structure in the $\phi$-photoproduction cross section
occurs lies very close to the threshold of $\Lambda$(1520)
production. Since the peak seems to have a rather narrow width, channel
coupling to the $\Lambda$(1520) resonance could be responsible for it.
In this paper we will present an analysis of these data in terms of the Groningen K-matrix
model which is based on an effective Lagrangian formulation for the reaction
kernel. This kernel obeys gauge invariance, crossing symmetry, and is covariant.
All these conditions are conserved in the K-matrix formulation which in addition
imposes the unitarity of the scattering matrix, provided that the kernel is hermitian.
The effective Lagrangian for the kernel is also consistent with chiral symmetry.
To investigate whether or not the observed structures in the $\phi$-photoproduction
cross section could be due to coupled-channel phenomena where
the $\Lambda$(1520) resonance is expected to play a crucial role,
we extend the Groningen
K-matrix approach to include the $\Lambda$(1520) resonance.
We will not consider, however, other channels such as
$K^* \Lambda$ and $K^* \Sigma$
since the $K^*$ meson has a larger width than the $\Lambda(1520)$ resonance
and the $\phi$-meson.

In \secref{kmatr} the basic principles of the Groningen
K-matrix approach will be shortly reviewed.
For the interaction kernels we follow previous
approaches where possible; for $\gamma N \to \phi N$ the Pomeron and meson exchange
model is employed~\cite{Titov:2003bk,Titov:2007fc},
and for $\gamma N \to K \Lambda(1520)$
the effective Lagrangian method is used~\cite{Nam:2005uq}.
The new ingredient introduced here is
the $K \Lambda(1520)\to \phi N$ coupling.
A brief description of the kernels is given in
Section  IIB -- D, while the effective Lagrangians used in these transition amplitudes are presented in Appendix A.

In the attempt to explain the peak structure in the cross section,
we pay attention to the following two aspects.
One is an on-shell kaon exchange effect in the transition kernel
$K\Lambda(1520) \to \phi N$, which is kinematically allowed and which produces
a singular structure in the energy region $E_\gamma \sim 2$ GeV.
We discuss this effect in \secref{on-shell-k}.
It turns out, however, that the inclusion of the two coupled channels
with the above singular behavior can not explain the observed peak structure in the
cross section.
As another possibility we introduce a nucleon resonance as a
bare pole in the coupled-channel approach in \secref{KKN}.
We assume that this nucleon resonance contains a large fraction
of hidden $s\bar{s}$ or $K \bar K$ components,
and that such a state will not be produced directly in photon-induced reactions and has
a sufficiently large coupling
to $K \Lambda(1520)$ as well as to $\phi N$.
In Section III, we present results for cross sections, including
the $t$-dependence, and spin observables.
The role of the on-shell kaon kinematics and nucleon resonances are
discussed in detail.
The final section is devoted to discussions and summary.

\section{Description of the Model}

Our model calculations are based on a coupled-channel calculation using the
K-matrix formulation as described in the following section.
We include $\pi N$, $\rho N$, $\eta N$, $K\Lambda$, $K\Sigma$, $K\Lambda(1520)$ and $\phi N$ channels.
The kernel for our coupled-channel calculation is derived from an effective Lagrangian method. Included
are the  $s$-, $t$-, and $u$-channel Born diagrams supplemented by the contact terms to ensure
gauge invariance where necessary. In addition a spectrum of low lying baryonic contributions is included in the $s$- and $u$-channels. A more complete account of the terms in the effective Lagrangians is given in Appendix A.

\subsection{K-matrix model}\seclab{kmatr}

The coupled-channel (or re-scattering) effects are included in our model
via the K-matrix formalism. In this section we present a short overview of
this approach where a more detailed description can be found in Refs.~\cite{Korchin:1998ff, Usov:2005wy, Scholten:2002tn}.

In the  K-matrix formalism the scattering matrix is written as
\begin{equation}\eqlab{T-matr}
  T = \frac{ K}{1-i K} \,.
\end{equation}
It is easy to check that the resulting scattering amplitude $S=1+2i T$ is
unitary provided that $K$ is Hermitian. The construction in \eqref{T-matr} can
be regarded as the re-summation of an infinite series of loop diagrams by
making a series expansion,
\begin{equation}
   T =  K + i K K + i^2 K K K + \cdots \,.
\end{equation}
The product of two $K$-matrices can be rewritten as a sum of different one-loop
contributions.
However, not the entire spectrum of loop contributions present in a systematic field-theoretical approach is generated in this way, and the missing ones should be accounted for in the kernel.
In constructing the kernel, care should be taken to avoid double
counting. For this reason we include in the kernel tree-level diagrams only
[Figs.~\ref{fig:feyn}(i)-\ref{fig:feyn}(iii)], modified with form-factors and
contact terms [Fig.~\ref{fig:feyn}(iv)]. The contact terms (or four-point
vertices) ensure gauge invariance of the model and express the model-dependence
which is introduced by
working with form factors (see~Appendix A). Contact terms and form factors can be
regarded as accounting for loop corrections which are not generated in the
K-matrix procedure, or for short-range effects which have been omitted from the
interaction Lagrangian. Inclusion of both $s$- and $u$-channel diagrams (Figs.~\ref{fig:feyn}(i) and \ref{fig:feyn}(ii), respectively) in the kernel insures the
compliance with crossing symmetry.

\begin{figure*}
\begin{center}
\includegraphics[width=0.5 \textwidth]{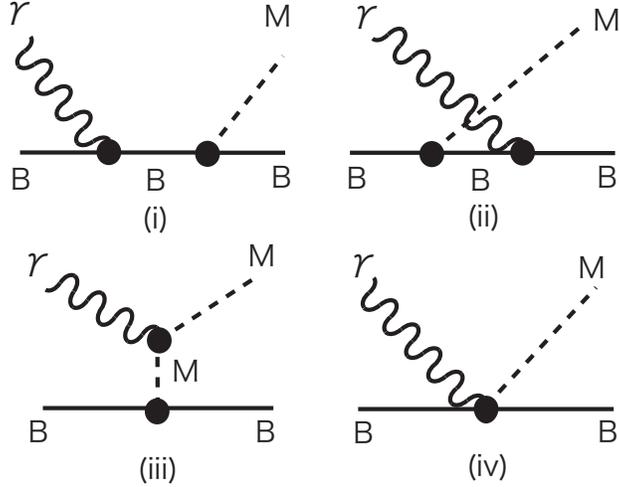}
\vskip -0.1in \caption{Feynman diagrams included in this work.
First row: $s$- and $u$-channel diagrams,
the external baryons $B$ in these diagrams are those included in the coupled channels,
$N, \ \Lambda, \ \Sigma, \ \Lambda(1520)$, and the intermediate ones can also be
$N^{*}, \ \Delta$ resonances in addition to those of the external ones when allowed.
$M$ stands for the mesons included in the model space
$\pi, \ \eta, \ K, \ \rho, \ \phi$.
Second row: $t$-channel contributions with meson exchange $\pi, \ K, \ \eta, \ K^{*}, \ \rho, \ \omega$, and the contact term required by the gauge invariance. \figlab{feyn}}
\end{center}
\end{figure*}

\subsection{The kernel $\gamma N \to \phi N$ }\seclab{phi-N}

Because of the OZI rule the nucleon pole ($s$- and $u$-type diagrams) contribution will be suppressed for the $\gamma N \to \phi N$ channel. The same will hold for the
nucleon resonance contributions if they are dominated by $u$ and $d$ quarks.
Therefore the $t$-channel contributions will be the most important in the kernel.
We include meson ($\pi, \eta$) exchange diagrams as well as the Pomeron
contribution.
Effective Lagrangians used in meson exchange amplitudes are presented in Appendix A.
As discussed in \secref{results}, the kernel is dominated by
the Pomeron contribution.
According to the recipe of Refs.~\cite{Titov:2003bk,Titov:2007fc, Donnachie:1987pu}, the amplitude of the Pomeron exchange is given by
\beq
\mathcal{M}
= \bar{u}(p_{N_{f}})\mathcal{M}_{\mu \nu}u(p_{N_{i}})\epsilon^{* \mu}_{\phi}\epsilon^{\nu}_{\gamma},
\eeq
where $\epsilon_{\phi}$ and $\epsilon_{\gamma}$ are the polarization vectors of
the $\phi$-meson and photon,
and $\mathcal{M}_{\mu \nu}$ is
\beq
\mathcal{M}^{\mu \nu}
= M(s,t)\Gamma^{\mu \nu}.
\eeq
Here the transition operator $\Gamma^{\mu \nu}$ reads
\begin{eqnarray*}
\Gamma^{\mu \nu}
&=& \Slash{k}_{\gamma}(g^{\mu \nu}-\frac{p^{\mu}_{\phi}p^{\nu}_{\phi}}{p^{2}_{\phi}})
-\gamma^{\nu}(k^{\mu}_{\gamma}-p^{\mu}_{\phi}\frac{k_{\gamma}\cdot p_{\phi}}{p^{2}_{\phi}}) \\
&-& (p^{\nu}_{\phi}-\frac{\bar{p}^{\nu}k_{\gamma}\cdot p_{\phi}}{\bar{p}\cdot k_{\gamma}})(\gamma^{\mu}-\frac{\Slash{p}_{\phi}p^{\mu}_{\phi}}{p^{2}_{\phi}}),
\end{eqnarray*}
with $\bar{p}=(p_{N_{i}}+p_{N_{f}})/2$. This amplitude satisfies gauge invariance. The factor $M(s,t)$ is written as
\beq
M(s,t)
&=& C_{p}F_{N}(t)F_{\phi}(t)\left(\frac{s}{s_{p}}\right)^{\alpha_{p}(t)-1}{\rm{exp}}
\left(-\frac{i\pi}{2}\alpha_{p}(t)\right),
\label{Mst}
\eeq
where $F_{N}(t)$ is the isoscalar form factor of the nucleon and $F_{\phi}(t)$ is the form factor for the photon-$\phi$ meson-Pomeron coupling.
They are parametrized as
\begin{eqnarray*}
F_{N}(t)
&=& \frac{4M^{2}_{N}-a^{2}_{N}t}{(4M^{2}_{N}-t)(1-t/t_{0})^{2}}, \\
F_{\phi}(t)
&=& \frac{2\mu^{2}_{0}}{(1-t/M_{\phi}^{2})(2\mu^{2}_{0}+M^{2}_{\phi}-t)}.
\end{eqnarray*}
In Eq.~(\ref{Mst}) the Pomeron trajectory $\alpha_{p}(t) = 1.08 + 0.25t$ is determined from hadron elastic scattering in the high energy region.
The strengh factor $C_{p}$ is
\beq
C_{p}
= \frac{6e\beta_{s}\beta_{u}}{\gamma_{\phi}},
\eeq
where $\gamma_{\phi} = 6.7$ is the $\phi$-meson decay constant. The constants $\beta_{s}$ and $\beta_{u}$ are Pomeron couplings with the strange quark
in a $\phi$-meson and the light quark in a proton, respectively.
For other parameters we use standard values for the Pomeron exchange model, that is, $t_{0}=0.7$ GeV$^2$, $\mu^{2}_{0} = 1.1$ GeV$^{2}$, $s_{P} = 4$ GeV$^{2}$, $a_{N}=2$, and $\beta_{s} = 1.44$ and $\beta_{u} = 2.04$ GeV$^{-1}$.

\subsection{The kernel $\gamma N \to K\Lambda(1520)$}

We can expect that the $\gamma N \to K\Lambda(1520)$ channel contributes to $\phi$-meson photoproduction through coupled-channel effects near the threshold region because the threshold of this channel is very close to that of the $\phi$N channel.
For $\Lambda(1520)$
photoproduction we use the model in Ref.~\cite{Nam:2005uq}, which is based on the
effective Lagrangian method. The relevant terms of the Lagrangian are presented in Appendix A.
In this model
the most important contribution is given by the contact term,
\beq
\mathcal{M}_{c}
&=& -\frac{eg_{KN\Lambda^{*}}}{m_{K}}\bar{u}^{\mu}\epsilon_{\mu}\gamma_{5}u\cdot F_{c},
\eeq
where $u^{\mu}$ and $\epsilon^{\mu}$ are the Rarita-Schwinger vector-spinor (see Appendix C) and photon polarization vector respectively, and
$F_{c}$ is a hadronic form factor (see Appendix A).
The coupling constant $g_{KN\Lambda^{*}}= -11$ ($\Lambda^{*}\equiv \Lambda(1520)$  is determined by the decay rate of $\Lambda(1520) \to KN$, $\Gamma_{\Lambda(1520) \to KN} \simeq 7$ MeV.
Gauge invariance is satisfied when all Born diagrams are summed.
This model successfully reproduces
the experimental data in the medium energy region ($\ 3 \le E_{\gamma} \le 5$ GeV).
Recent LEPS data~\cite{Muramatsu:2009zp} also support this model in the low energy region.

\subsection{The kernel $K\Lambda(1520) \to \phi N$}

For the terms in the effective interaction Lagrangian for the $K\Lambda(1520) \to \phi N$ kernel, which plays a crucial
role in coupled-channel effects, we choose
\beq
\mathcal{L}_{KN\Lambda^{*}}
&=& i\frac{g_{KN\Lambda^{*}}}{m_{K} M_{\Lambda^*} } \bar{N}\gamma_{5}
 \Big[ (\partial_{\alpha}K^{+})(\s{\partial}\Lambda^{*\alpha}) -
 (\partial_{\mu}K^{+}) \gamma_\alpha (\partial^\mu \Lambda^{*\alpha}) \Big] \;, \eqlab{KNL*} \\
\mathcal{L}_{\phi KK}
&=& -ig_{\phi}(\partial^{\mu}K^{-}K^{+}-\partial^{\mu}K^{+}K^{-})\phi_{\mu}, \\
\mathcal{L}_{\phi \Lambda^{*}\Lambda^{*}}
&=& - {g_{\phi} \over M_{\Lambda^*}^2} (\partial^\rho \bar{\Lambda}^{*\beta})
 \Big[ g_{\alpha\beta} g_{\rho\tau} - g_{\tau\beta} g_{\rho\alpha} \Big]
 \Big( -\gamma_{\mu}+\frac{\kappa_{\phi}}{2M_{\Lambda^{*}}}\sigma_{\mu
 \nu}\partial^{\nu} \Big) \phi^{\mu}
 (\partial^\tau \Lambda^{*\alpha} ) \;,
\eeq
where we assume universality for $\phi$-meson and hadron couplings including strangeness, that is, $g_{\phi}=g_{\phi KK}$.
The vertices are chosen such that the coupling to the spin-$1/2$-component of the intermediate Rarita-Schwinger
propagator vanishes (so-called gauge-invariant vertices).  From the decay width $\Gamma_{\phi \to
KK}$ we obtain the value $|g_{\phi}|=4.7$.
For simplicity, we take $\kappa_{\phi}=0$.
Here we employ these Lagrangians rather than those similar to (A5, A6)
with the photon field $A_{\mu}$ replaced by $\phi_{\mu}$ since for coupling to
the photon field the magnetic coupling (set to zero for $\phi$-meson production)
is dominant together with the associated contact term.
For coupled-channel calculations a proper treatment of off-shell properties
is important and the choice of (8)-(10) is one way to reduce
the ambiguities in the off-shell propagator of a spin $3/2$-particle.
Replacing the derivative in \eqref{KNL*} by a covariant
derivative, that is, $\partial_{\mu}K^{+} \to (\partial_{\mu} -
ig_{\phi}\phi_{\mu})K^{+}$ (the minus sign is due to $Q_{s}=-1$ in the case of $K^{+}$)
and $\partial_{\mu}\Lambda^{*} \to (\partial_{\mu} + ig_{\phi}\phi_{\mu})\Lambda^{*}$
(the plus sign is due to $Q_{s}=+1 $ in the case of $\Lambda^{*}$),
we obtain the contact term
\beq
\mathcal{L}_{\phi K N \Lambda^{*}}
&=& -\frac{g_{\phi}g_{KN\Lambda^{*}}}{m_{K}M_{\Lambda^{*}}}\bar{N}\gamma_{5}
\Big[ -\phi_{\alpha}K^{+}\Slash{\partial}\Lambda^{* \alpha}+\partial_{\alpha}K^{+}\gamma_{\mu}\phi^{\mu}\Lambda^{* \alpha} \Big] \;.
\eeq
Here we find an additional term, the second term of (11) which does not exist in (A7) for
on-shell $\Lambda^{*}$. This term, however, is shown to be suppressed by the order
$O(p/M)$ and has a negligible contribution to the scattering amplitude.
Using these effective interaction Lagrangians we obtain the transition amplitudes $K\Lambda(1520) \to \phi N$
\beq
\mathcal{M}_{u}
&=& i\frac{g_{\phi}g_{KN\Lambda^{*}}}{m_{K}M^{3}_{\Lambda^{*}}}\bar{u}(p_{N})\gamma_{5}[p_{K \nu}\Slash{q}_{u}-(p_{K}\cdot q_{u})\gamma_{\nu}]D^{\nu \beta}_{3/2}[g_{\alpha \beta}
(q_{u}\cdot p_{\Lambda^{*}})-p_{\Lambda^{*} \beta}q_{u \alpha}]\Slash{\epsilon}_{\phi}u^{\alpha}(p_{\Lambda^{* }}), \\
\label{u-channel}
\mathcal{M}_{t}
&=& -\frac{g_{\phi}g_{KN\Lambda^{*}}}{m_{K}}\frac{1}{q^{2}_{t}-m_{K}^{2}}
 (q_{t}-p_{K})\cdot \epsilon_{\phi}\bar{u}(p_{N})\gamma_{5}(q_{t}\cdot u(p_{\Lambda^{*}})), \\
\mathcal{M}_{c} &=&\frac{g_{\phi}g_{KN\Lambda^{*}}}{m_{K}M_{\Lambda^{*}}}\bar{u}(p_{N})\gamma_{5}
(\Slash{p}_{\Lambda^{*}}\epsilon_{\phi \alpha} - \Slash{\epsilon}_{\phi}p_{K \alpha}) u^{\alpha}(p_{\Lambda^{*}}),
\eeq
where the four-momenta of the kaon, $\Lambda^{*}$, $\phi$, $N$ are denoted by $p_{K}$,
$p_{\Lambda^{*}}$, $p_{\phi}$ and $p_{N}$ respectively, and we have defined
$q_{t}=p_{\phi}-p_{K}$. $\epsilon^{\mu}_{\phi}$ is the polarization vector of the
$\phi$-meson. $D^{\alpha \beta}_{3/2}$ stands for the spin-$3/2$ propagator
\beq
D^{\alpha \beta}_{3/2}(q_{u})
&=& -i\frac{\Slash{q}_{u}+M_{\Lambda^{*}}}{q^{2}-M_{\Lambda^{*}}^{2}}
 \Big[ g^{\alpha \beta}-\frac{1}{3}\gamma^{\alpha}\gamma^{\beta}-
 \frac{2}{3M_{\Lambda^{*}}^{2}}q_{u}^{\alpha}q_{u}^{\beta}
 -\frac{q_{u}^{\alpha}\gamma^{\beta}-q_{u}^{\beta}\gamma^{\alpha}}{3M_{\Lambda^{*}}} \Big],
\label{3/2prop}
\eeq
where $q_u=p_{\Lambda^{*}}-p_{\phi}$. Assuming $g_{\phi NN}=0$, as suggested by
the OZI rule, the $s$-channel amplitude vanishes.

\subsection{On-shell kaon exchange}\seclab{on-shell-k}

\begin{figure*}
\begin{center}
\includegraphics[width=0.3 \textwidth]{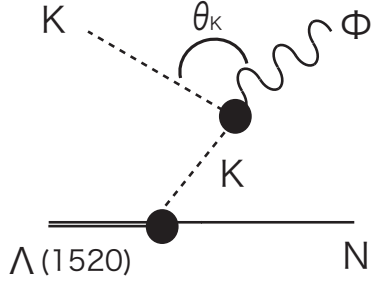}
\vskip -0.1in \caption{The $t$-channel diagram in the kernel $K\Lambda(1520) \to \phi N$.} \label{t-channel_d}
\figlab{t-diagram}
\end{center}
\end{figure*}

In the $t$-channel contribution to the $K\Lambda(1520) \to \phi N$ kernel, (Fig.~\ref{t-channel_d}),
the exchanged kaon can reach the on-shell pole for certain
kinematical conditions. The on-shell condition is
\beq
t(W(E_{\gamma}), {\rm{cos}}\theta_{K})-m_{K}^{2}=0, \eqlab{on-shell}
\eeq
where $t$ is the momentum transfer as a function of the invariant mass ($W$),
which in turn depends on the energy of the in-coming photon
in the $\gamma N \to \phi N$ channel and
on ${\rm{cos}}\theta_{K}$, where $\theta_{K}$ is the angle between the out-going $\phi$-meson
and the in-coming kaon as shown in \figref{t-diagram}.
Therefore, \eqref{on-shell} gives a relation between $E_{\gamma}$ and ${\rm{cos}}\theta_{K}$ which is shown by the solid line in the l.h.s.\ of \figref{solution}.
Since $|{\rm{cos}}\theta_{K}| \leq 1$ the solution of \eqref{on-shell} is
kinematically limited to a narrow energy region given by the photon energy $1.7 \ {\rm{GeV}}
\lsim E_{\gamma} \lsim 2.1 \ {\rm{GeV}}$.
In the l.h.s.\ of \figref{solution} the allowed region is colored in blue.
An interesting point is that the blue region $E_{\gamma} \lsim 2.1$ GeV corresponds to
the region where the peak structure in the production cross section is observed.

The fact that the intermediate kaon may become on-shell is similar to the threshold effect
in the coupled-channel dynamics, and might be responsible for a singular behavior
in the kinematically allowed region.
\begin{figure*}
\begin{tabular}{cc}
\begin{minipage}{0.5\hsize}
\begin{center}
\includegraphics[width=0.7 \textwidth]{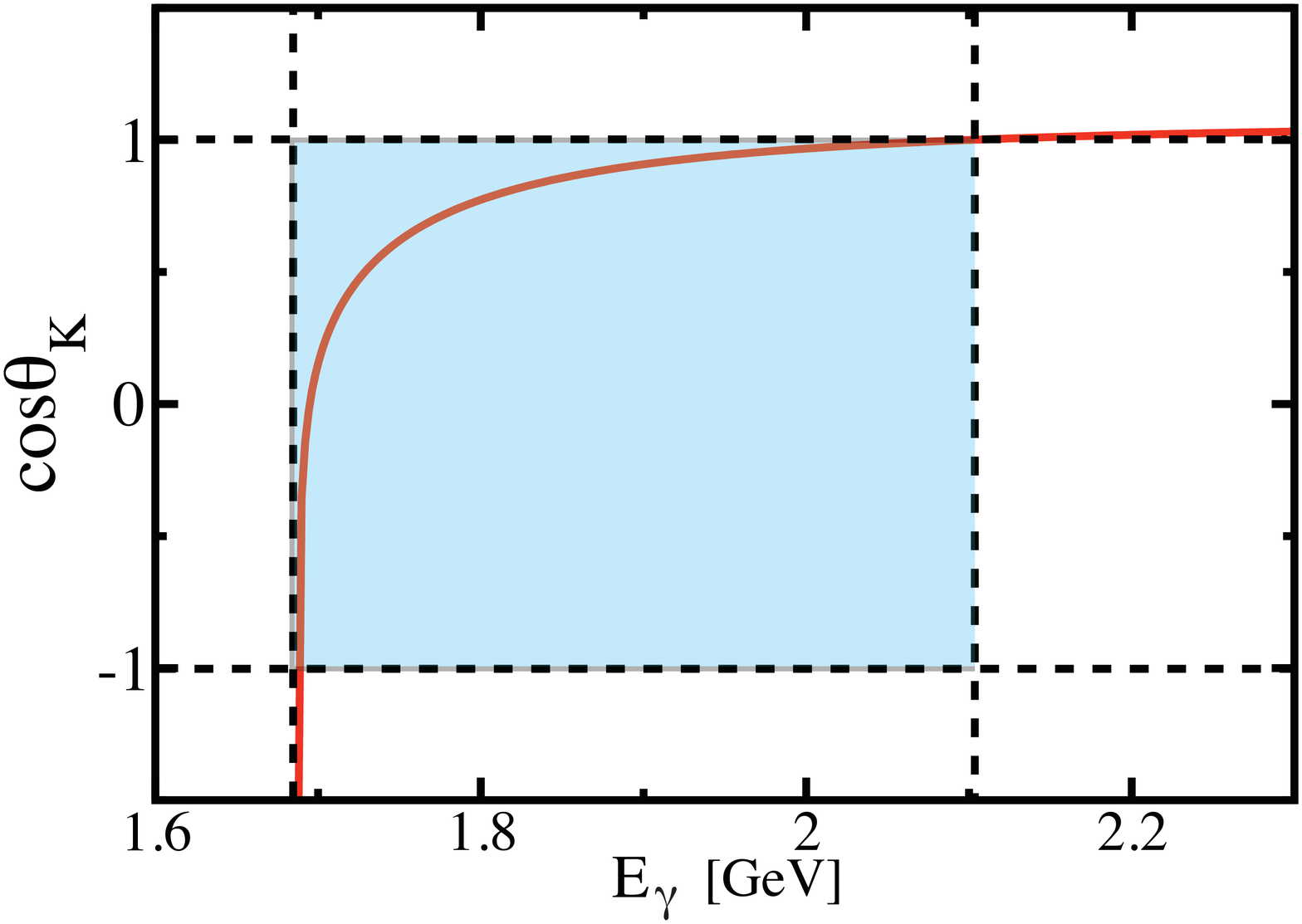}
\vskip -0.1in
\end{center}
\end{minipage}
\begin{minipage}{0.5\hsize}
\begin{center}
\includegraphics[width=0.68 \textwidth]{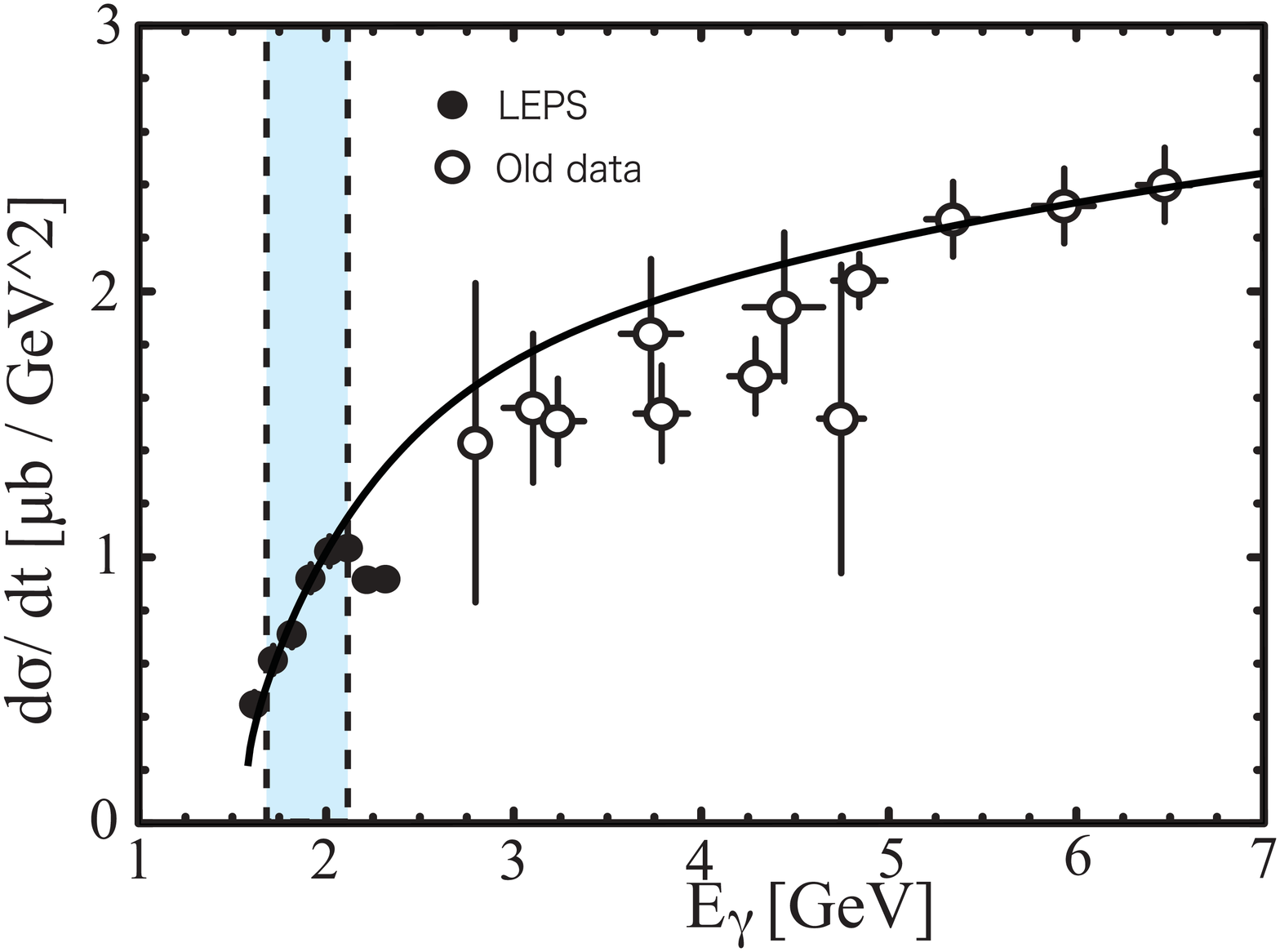}
\vskip -0.1in
\end{center}
\end{minipage}
\end{tabular}
\caption{[color online]
The solution to the on-shell condition of the intermediate kaon (l.h.s.). The kinematically allowed region,
$-1 \leq {\rm{cos}}\theta_{K} \leq 1$ or equivalently $1.7$ GeV $\leq E_{\gamma} \leq 2.1$ GeV is colored blue with
the boundary shown by the dashed lines. The allowed region in $E_{\gamma}$ is also shown on the r.h.s.\ in the plot of
the cross section.
} \figlab{solution}
\end{figure*}
In reality this pole contribution is washed out due to the fact that
the $\Lambda(1520)$ resonance as well as the $\phi$-meson are unstable particles. If they were
stable, they could not couple to the on-shell $K\bar{K}N$ intermediate state.
The resulting small but finite width of these particles is however
not yet taken into account in the calculations. This width can
be folded with the divergent pole contribution, creating a resonance-like
structure for the matrix element. To simulate this we introduce an effective
width for the exchanged kaon reflecting the decay width of the external
$\phi$-meson and the $\Lambda(1520)$ resonance,
\beq
\frac{i}{t-m_{K}^{2}} &\to& \frac{i}{t-m_{K}^{2}-im_{K}\Gamma_{K}}.
\eqlab{ficwidth}
\eeq

In Appendix B we give an estimate for the value of the effective width for the intermediate
kaon. We have obtained a value for $\Gamma_{K}$ of
\beq
\Gamma_{K} > 7.79 \ [\mbox{MeV}] ,
\eeq
in the on-shell region (see Appendix B).
In the calculations we have chosen the value $\Gamma_{K}=10$ MeV.
In \secref{results} we will discuss
the dependence of the calculated $\phi$-photoproduction cross section on
$\Gamma_{K}$ which is rather small.

\subsection{A nucleon resonance with large $s\bar{s}$ components}\seclab{KKN}

In the present discussions we have not yet considered any nucleon resonance
contributions. A normal nucleon resonance will not contribute to the $K\Lambda(1520) \to \phi
N$ kernel because of the OZI rule, however, if
the resonance contains a hidden $s\bar{s}$ component, it may strongly couple to
$\phi N$ and/or $K\Lambda(1520)$ states and will contribute
through the $s$-channel process.
In this paper, as one possibility, we introduce such a nucleon resonance as a pole
term since in the K-matrix approach resonances are not generated dynamically.
This pole term, with the bare mass $M_{N^{*}}$, is introduced in the kernels of
$K\Lambda(1520) \to K\Lambda(1520)$,
$K\Lambda(1520) \to \phi N$ and $\phi N \to \phi N$.
This resonance can thus be regarded as a $\phi N$, $K\Lambda(1520)$, or a $K\bar{K}N$
quasi-bound state or a superposition of such states.
The possibility of such a resonance is discussed for slightly lower energy region in Refs.~\cite{KanadaEn'yo:2008wm,Jido:2008kp}. In contrast to the mass,
its finite decay width is dynamically generated through the coupling to
the $K\Lambda(1520)$ and $\phi N$ channels.
Furthermore, we assume that the resonance is not directly excited in
photoproduction. In spite of this, it may contribute to $\phi$-mson
photoproduction through coupled-channels effects as investigated in this work.
Since we do not know the spin and parity of the resonance,
we assume them $J^{P}=1/2^{\pm}$, for simplicity.
The kernels in which the nucleon resonance appear in the $s$-channel
are given by
\beq
\mathcal{M}^{N^{*}}_{\phi N \to K\Lambda^{*}}
&=&-i\frac{g_{K\Lambda^{*}N^{*}}}{m_{K}}\frac{g_{\phi}\kappa_{\phi N N^{*}}}{4M_{N}}\bar{u}(p_{N})[\Slash{\epsilon}_{\phi}\Slash{p}_{\phi}-\Slash{p}_{\phi}\Slash{\epsilon}_{\phi}]\Gamma_{1}
\frac{i}{\Slash{p}_{N^{*}}-M_{N^{*}}}\Gamma_{2}p_{K \mu} u^{\mu}(p_{\Lambda^{*}}), \\
\mathcal{M}^{N^{*}}_{K\Lambda^{*} \to K\Lambda^{*}}
&=&i\left(\frac{g_{K\Lambda^{*}N^{*}}}{m_{K}}\right)^{2}\bar{u}^{\mu}(p_{\Lambda^{*}}^{'})
p_{K \mu}^{'}\Gamma_{2}\frac{i}{\Slash{p}_{N^{*}}-M_{N^{*}}}\Gamma_{2}p_{K \nu} u^{\nu}(p_{\Lambda^{*}}), \\
\mathcal{M}^{N^{*}}_{\phi N \to \phi N}
&=& \left(\frac{g_{\phi}\kappa_{\phi NN^{*}}}{4M_{N}}\right)^{2}\bar{u}(p_{N}^{'})
[\Slash{\epsilon}_{\phi}^{'}\Slash{p}_{\phi}^{'}-\Slash{\epsilon}_{\phi}^{'}\Slash{p}_{\phi}^{'}]\Gamma_{1}
\frac{i}{\Slash{p}_{N^{*}}-M_{N^{*}}}\Gamma_{1}
[\Slash{\epsilon}_{\phi}\Slash{p}_{\phi}-\Slash{\epsilon}_{\phi}\Slash{p}_{\phi}]u(p_{N}),
\eeq
where $g_{K\Lambda^{*} N^{*}}$, $\kappa_{\phi N N^{*}}$ and $M_{N^{*}}$ are free parameters
which characterize the $N^{*}$ resonance. $\Gamma_{1}$ ($\Gamma_{2}$) equals
$1_{4\times4}$ ($\gamma_{5}$) for the positive-parity resonance and equals
$\gamma_{5}$ ($1_{4\times4}$) for negative-parity resonance, respectively.

In the following section we investigate the extent to which
such an $N^{*}$ resonance with
hidden strangeness can be responsible for the peak structure observed in
$\phi$-photoproduction.

\section{Results and Discussions}\seclab{results}

In this section we present our numerical results for the reaction cross section of $\gamma N \to \phi N$.
To investigate the roles of coupled channels, we first discuss the contributions of the different channels.
Next we introduce the
$N^{*}$ resonance with hidden-strangeness in our model. Finally we show our results for the
$t$-dependence of the cross section and spin-density matrices.

\subsection{The kernel $\gamma N \to K \Lambda(1520)$}

As was mentioned before, one of coupled-channel contributions to $\phi$-meson
photoproduction which
is potentially interesting is the one going via the $K \Lambda(1520)$
intermediate channel. The reason is that the threshold for this channel lies
close to that of $\phi N$. The magnitude of this coupled-channel
contribution is determined by the product of kernels for
$\gamma N \to K \Lambda(1520)$ and $K \Lambda(1520) \to \phi N$.

To fix the magnitude of $\gamma N \to K \Lambda(1520)$ kernel
we compare the calculated
cross section for this reaction with data in \figref{KLst_tot}. The calculated
cross sections depend strongly on the cut-off parameter in the hadronic form
factor (see Appendix A). Following Ref.~\cite{Nam:2005uq} a good agreement with the data in the medium energy region ($3000 \ {\rm{MeV}} \le E_{\gamma} \le 5000$ MeV) is obtained using a value of $\Lambda_{c} = 0.7$~GeV.
At higher energies the calculation overestimates the data, however,
this is outside the region of interest for the present investigation.
Care should be taken with using the parameters of Ref.~\cite{Nam:2005uq} for
the reaction of $\gamma N \to K\Lambda(1520)$ since coupled-channels effects will
contribute.
Our strategy is first to fix each kernel to reproduce the corresponding data if available and to introduce
the coupled-channel effects afterwards.
If the latter effect is important, we will reconsider the kernels themselves such that the coupled-channel results will
reproduce the data.
For the $K\Lambda(1520)$ photo-production, we have found that the coupled-effects are not very important.

\begin{figure*}
\begin{center}
\includegraphics[width=0.5 \textwidth]{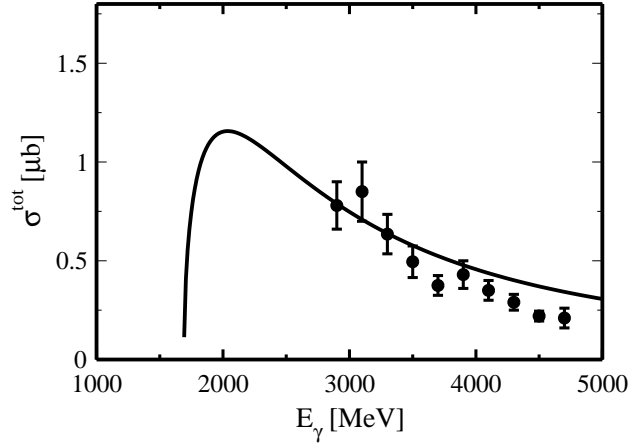}
\vskip -0.1in \caption{The total cross section for $\gamma p \to
K^{+} \Lambda(1520)$ as function of the photon energy in the laboratory system $E_\gamma$, calculated with a cut-off of $\Lambda_{c}=0.7$~GeV. The data are
taken from Ref.~\cite{Barber:1980zv}.} \figlab{KLst_tot}
\end{center}
\end{figure*}

\subsection{The kernel $K\Lambda(1520) \to \phi N$}

The $K\Lambda(1520) \to \phi N$ kernel is one of the most important ingredients in the present coupled-channel analysis.  In \figref{on_shell}, various contributions to the differential cross section of the $\phi$-photoproduction,
$d\sigma/dt (\theta = 0)$ are shown when only $K\Lambda(1520) \to \phi N$ term is included in the kernel to highlight the effect of the coupled channels of $K\Lambda(1520)$ and $\phi N$.
In this case, for example, the lowest order process is given by $\gamma N \to K\Lambda(1520) \to \phi N$.

An important aspect of the kernel for the $K\Lambda(1520) \to \phi N$ channel
is the pole in the $t$-channel kaon that can be hit for certain kinematics. In
\secref{on-shell-k} we argued that this singularity could be treated effectively by
assigning a finite width $\Gamma_{K}$ to the intermediate kaon propagator, where
we related the width to
the physical decay widths of the $\Lambda(1520)$ resonance and $\phi$-meson
in the in-coming and out-going channels.
To test the importance of the width attributed to the intermediate kaon
we have calculated the cross section for different values of $\Gamma_{K}$.
\begin{figure*}
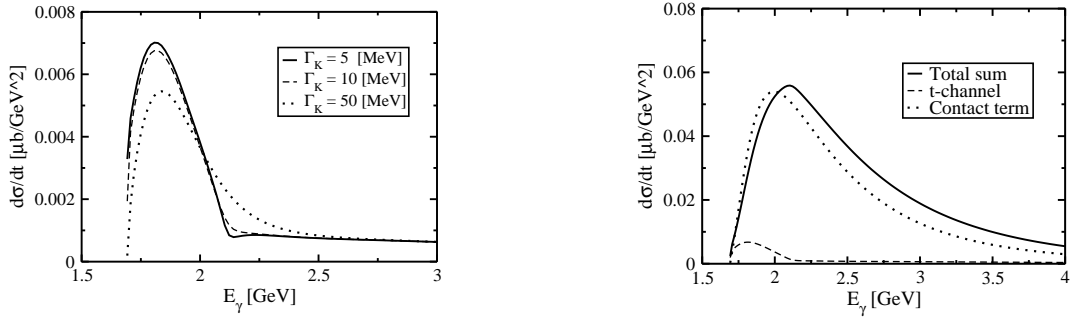

\begin{tabular}{cc}
\begin{minipage}{0.5\hsize}
\begin{center}
\includegraphics[width=0.7 \textwidth]{only_KLst_t-channel.eps}
\end{center}
\end{minipage}
\begin{minipage}{0.5\hsize}
\begin{center}
\includegraphics[width=0.7 \textwidth]{only_KLst_all-channel.eps}
\end{center}
\end{minipage}
\end{tabular}
\caption{The differential cross section of $\gamma N \to \phi N$ at zero degrees,
$d\sigma/dt (\theta=0)$. The l.h.s.\ shows the effects of
changing the width of the intermediate kaon, $\Gamma_{K} = 5$~MeV (solid line),
$\Gamma_{K} = 10$~MeV (dashed line), and $\Gamma_{K} = 50$~MeV (dotted line),
in a calculation in which only the $t$-channel kaon exchange contribution is included for the
$K\Lambda(1520) \to \phi N$ matrix elements.
The r.h.s.\ shows the result including the full
matrix element (solid line), only the contact term (dotted line) and only the $t$-channel
kaon exchange contribution (dashed line).}
\figlab{on_shell}
\end{figure*}
In \figref{on_shell} the l.h.s.\ of the figure shows $d\sigma/dt(\theta=0)$ when
only the $t$-channel contribution of kaon exchange is included in the kernel
$K\Lambda(1520) \to \phi N$.
Different lines correspond to those calculated with different values of $\Gamma_{K}$,
$\Gamma_{K}=5$ MeV (solid line), $10$ MeV (dashed line) and $50$ MeV (dotted line).
Variation of $\Gamma_{K}$ does not influence on the result very much
as shown in the figure.
Based on the decay width of the $\phi$-meson and that of the $\Lambda(1520)$ resonance one expects $\Gamma_{K} \approx 10$~MeV as argued in the previous section.
Therefore, we will use this value in the following calculations.

Interestingly, the cross section shown in the l.h.s.\ of \figref{on_shell} shows a peak structure
in the same energy region as the experimental data.
However the magnitude of the peak in the calculation is very small.
The effects of the other contributions of the $K\Lambda(1520) \to \phi N$ kernel from the
$u$-channel and contact term is investigated in the r.h.s.\ of \figref{on_shell} where we can see that the
dominant contribution is generated by the contact term as shown by the dotted line which is compared with the solid line for the case including all terms.
This is similar in structure as that in the $\gamma N \to K\Lambda(1520)$ channel~\cite{Nam:2005uq}.
The $t$-channel kaon exchange
contribution is thus buried under the contribution from the contact term (dotted line). The $u$-channel
contribution is also negligibly small.

\begin{figure*}
\begin{center}
\includegraphics[width=0.5 \textwidth]{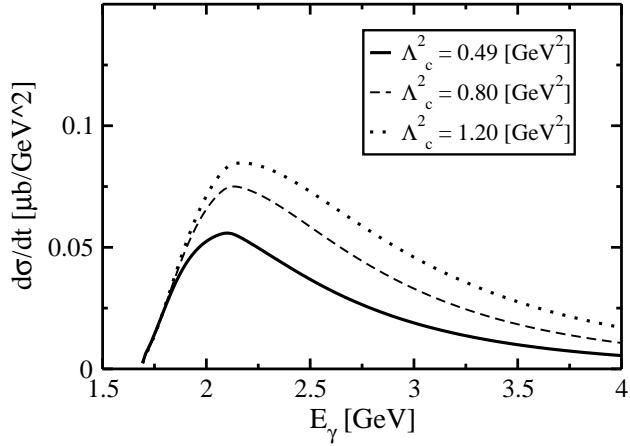}
\end{center}
\caption{Differential cross sections for $\phi$-meson
photoproduction at zero degrees for different values of cut-off parameters in
the kernel for $K\Lambda(1520) \to \phi N$,  $\Lambda_{c}^2 = 0.49, \ 0.80$, and
1.20 GeV$^2$ for the solid, dashed, and dotted lines, respectively.} \figlab{CC-effect}
\end{figure*}

We have also investigated the cut-off dependence of the kernel.
The solid line is for $\Lambda^{2}_{c}=0.49 \ {\rm{GeV}}^{2}$ ($\Lambda_{c}=0.7$ GeV) as corresponding to the solid line in the r.h.s.\ of \figref{on_shell}, the dashed line for
$\Lambda^{2}_{c} = 0.8 \ {\rm{GeV}}^{2}$, the dotted line for
$\Lambda^{2}_{c} = 1.2 \ {\rm{GeV}}^{2}$ which is the value used for other kernels
such $\gamma N \to K\Lambda(1116)$, $\gamma N \to K\Sigma(1193)$ etc.
From the figure we see that the cut-off dependence is not very strong.

One important conclusion that should be drawn from the results presented here is that the cross section calculated by including only the Born terms in the kernel $K\Lambda(1520) \to \phi N$ is not very large by itself, typically,
$d\sigma/dt (\theta=0) \leq 0.1 \ \mu{\rm{b/GeV^{2}}}$ which is smaller than the experimental values of order $1 \ \mu{\rm{b/GeV^{2}}}$.
However, when other dominant terms especially from the Pomeron exchange term is included, the interference among those terms becomes
sizable as we will see in the following sections.

\subsection{Differential cross section at $\theta = 0$}

The dominant contribution to the $\phi$-photoproduction cross section is
derived from the Pomeron exchange diagram. The magnitude of the Pomeron
contribution is fixed by the measured cross section for $E_\gamma \gsim 3$~GeV. As shown in Ref.~\cite{Titov:2007fc},
other hadronic contibutions such as $\sigma$, $\pi$ and $\eta$ exchange diagrams are orders of
magnitude smaller at higher photon energies.

\begin{figure*}
\begin{center}
\includegraphics[width=0.5 \textwidth]{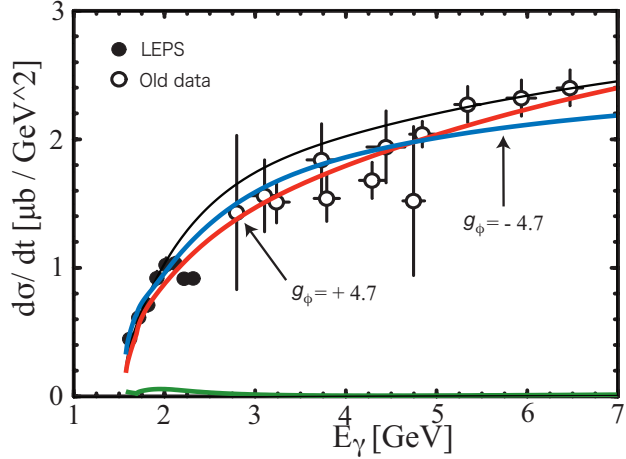}
\vskip -0.1in
\end{center}
\caption{[color online] The differential $\phi$-meson photoproduction cross
section at zero degrees. The black curve shows the results
of the tree level calculation, including the Pomeron contribution. The red (blue)curve
shows the full coupled-channel result with $g_{\phi} = +4.7$ ($g_{\phi} = -4.7$).
The green line is the effect of only coupled channel effects not only
K$\Lambda(1520)$ channel but also others.} \figlab{differ}
\end{figure*}

In our model, the value of the cut-off parameter in the hadronic form factor is
treated as a free parameter as well as the sign of $g_{\phi}$. The hadronic
form factor is discussed in more detail in Appendix A. We have calculated the
differential cross section for the two cases $g_{\phi} = + 4.7$ and
$g_{\phi} =- 4.7$. The absolute value is determined by the $\phi$-meson decay width.
In \figref{differ} we show the differential cross section calculated in our coupled-channel approach at $\theta = 0$, $d\sigma/dt(\theta=0)$
as a function of the photon energy in the laboratory fame $E_{\gamma}$.
As explained previously, we have included as coupled channels not only the $\Lambda(1520)$ but also those containing
the ground state hyperons.
As we have anticipated, the main contribution is due to  the Pomeron exchange.
The most important coupled-channel effect is mainly from the $K\Lambda(1520)$, where the interference of the amplitude of the Pomeron exchange and the coupled channel $K\Lambda(1520)$ is sizable.
Other coupled-channel effects, i.e., those from the ground state hyperons are not significant.
In \figref{differ}, the thin solid line is for the result of the Pomeron exchange without the coupled channels and the other two thick lines for those with the coupled channels;
the red line for positive $g_{\phi} = +4.7$ and the blue line for negative $g_{\phi} = -4.7$.
However as mentioned
in the previous section,
the on-shell kaon effect does not reproduce the peak structure with the sufficient strength near the threshold region.

The Pomeron contribution whose strength is determined by the cross sections beyond $E_{\gamma} \geq 3$ GeV can
be smoothly extrapolated downward the threshold energy region, where it traces the peak value at $E_{\gamma} \approx 2$ GeV.
In this case, the structure at around $E_{\gamma} \approx 2$ GeV can rather be interpreted as a dip at around $E_{\gamma} \approx 2.3$ GeV.
The structure of either peak or dip can be realized by different phases of  amplitudes
through interference.
We have partly tested this already by changing the sign of $g_{\phi}$, which changes the sign of the kernel of such as
$K\Lambda(1520) \to \phi N$, $K\Lambda(1116) \to \phi N$ and $K\Sigma(1193) \to \phi N$.
As shown by the red and blue lines in \figref{differ}, we could not reproduce the peak/dip structure by the interference
in the coupled-channel approach.

\subsection{Introduction of the $N^{*}$ resonance with large $s\bar{s}$ components}

In the previous section, we have observed that the peak at around $2$ GeV looks rather a dip around $2.3$ GeV after a global fit of the Pomeron exchange amplitude to the experimental data.
For this reason we refer to it as the dip structure rather than the peak structure in the following discussions.
In order to reproduce the observed dip structure in the
$\phi$-meson photoproduction cross section at forward angle we will introduce in this section an $N^{*}$
resonance with large $s\bar{s}$ components as discussed in \secref{KKN}.
We introduce the resonance as a pole term of mass $M_{N^{*}}$.
The decay width is then generated dynamically after the coupled channel equation is solved, and is related to the coupling strength of the pole term to the $\phi N$ and $K \Lambda(1520)$ channels.
The strengths are treated as parameters in order to reproduce the depth and the width of the dip.
For simplicity we have limited ourselves only to a spin-$1/2$
resonance.
We found that an excellent fit to the data could be obtained by assuming a negative parity resonance,
much better so than for a positive resonance.
This is due to the interference of the partial waves naturally entering in the calculation with those of the $N^{*}$
resonance.
For instance the relative motion of the $\phi N$ channel which is coupled to the $J^{P} = 1/2^{-}$ partial wave can be S-wave while for a $J^{P} = 1/2^{+}$ it should be P-wave.
In the present treatment of the resonance, where the coupling to the channels generates the decay width, the magnitude
of the width, energy dependence and the resulting interference pattern with the other amplitude depend on the spin and parity of the resonance in a complicated manner.
In the following we will present results for a resonance with $J^{P} = 1/2^{-}$.
For coupling parameters we have adopted the
values as given in Table~\ref{reson_models}
where two parameter sets, A and B, are given, corresponding to two signs for $g_{\phi}$,
$g_{\phi} = +4.7$ (set A) and $g_{\phi} = - 4.7$ (set B).
In \figref{reson_differ} we show
the effect of including a spin-$1/2^-$ $N^{*}$ resonance on
$d\sigma/dt (\theta = 0)$.
The set A and B can both reproduce the
dip structure through a destructive interference.
The central point
of the dip corresponds roughly to the pole position of the resonance
when the width of the resonance is not too wide.
The width of the resonance can be estimated from the extension of the dip in the cross section shown in
\figref{reson_differ} to be around $100$~MeV for both parameter sets A and B.
An experiment with photon energies $E_{\gamma} \sim 2.25$~GeV and beyond will be able to
provide more information on resonance parameters.

\begin{table}[h]
\begin{tabular}{|c||c|c|c|c|}
\hline
Sets & $g_{\phi}$ & $g_{K\Lambda(1520)R}$ & $\kappa_{\phi NR}$ & $M_{R}$ (GeV) \\
\hline
Set A &$+ 4.7$ & $-4.24$ & $0.07$ & $2.25$ \\ \hline
Set B &$- 4.7$ & $-4.90$ & $0.06$ & $2.26$ \\
\hline
\end{tabular}
\caption{ Resonance parameters}
\label{reson_models}
\end{table}

\begin{figure*}
\begin{tabular}{cc}
\begin{minipage}{0.5\hsize}
\begin{center}
\includegraphics[width=0.7 \textwidth]{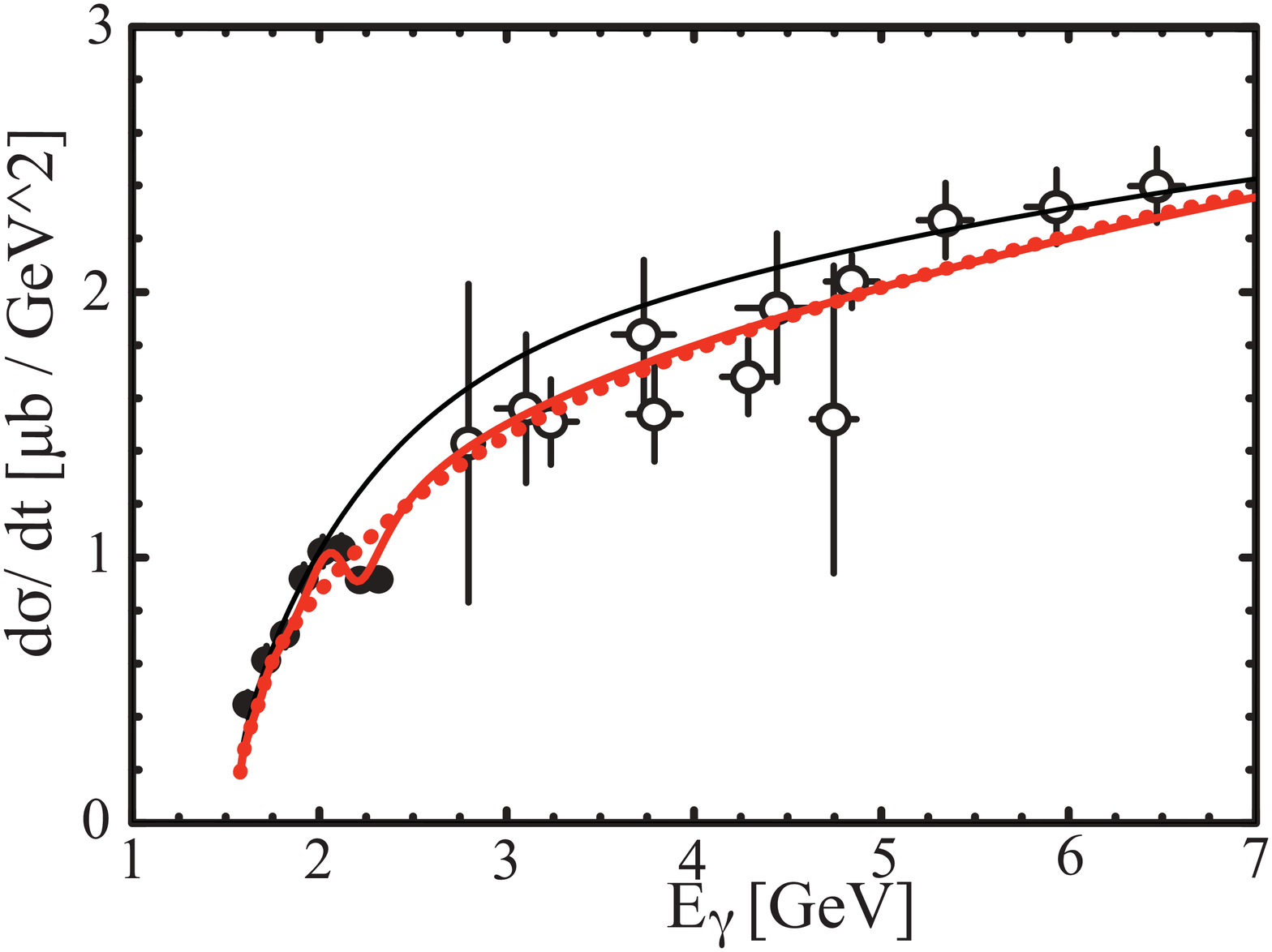}
\end{center}
\end{minipage}
\begin{minipage}{0.5\hsize}
\begin{center}
\includegraphics[width=0.7 \textwidth]{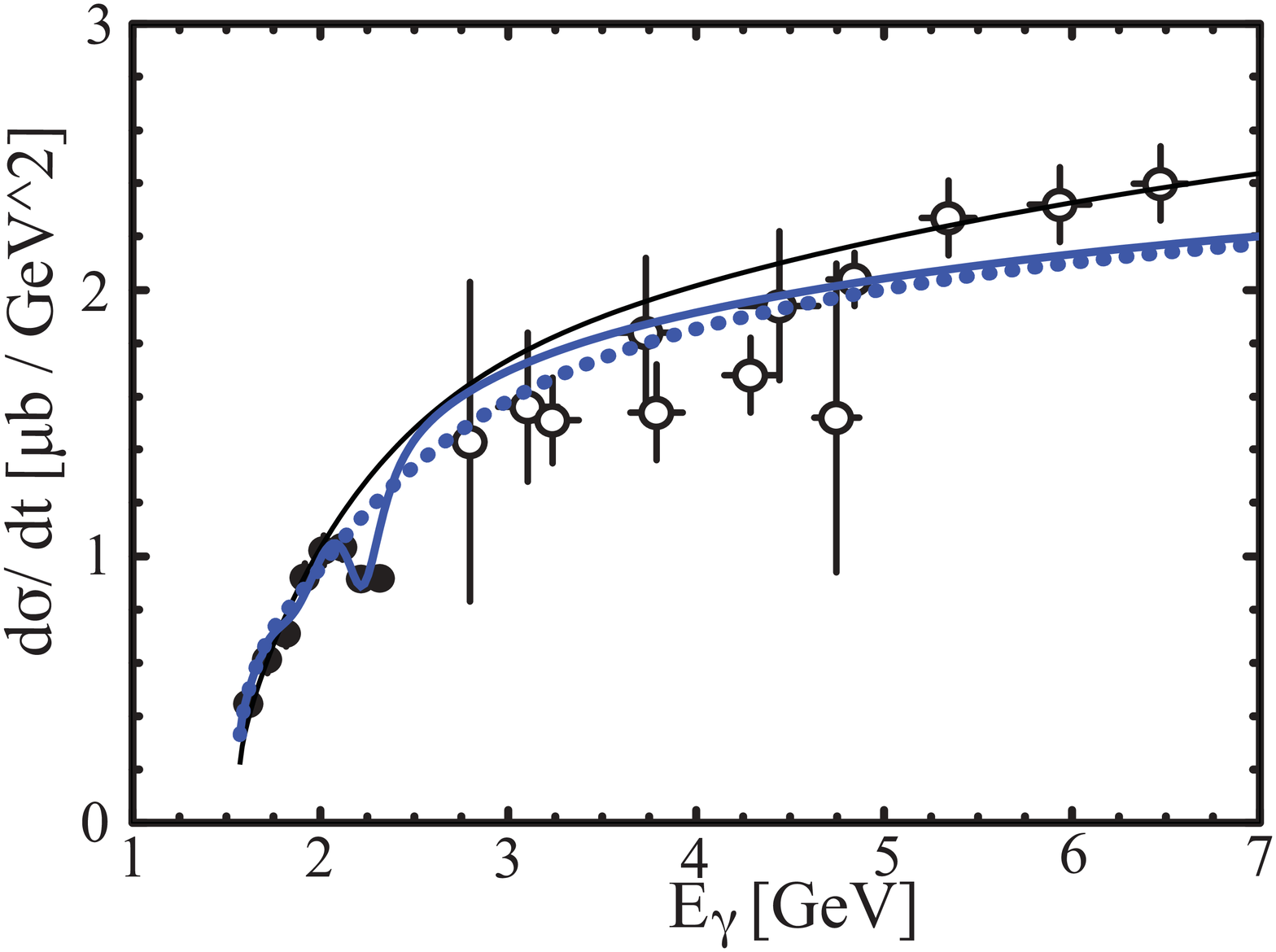}
\end{center}
\end{minipage}
\end{tabular}
\caption{[color online] The effect of a spin-$1/2^-$ $N^{*}$ resonance on the
forward angle $\phi$-meson photoproduction cross section. The thin black line in the both figures are tree level results.
In the l.h.s.\ (r.h.s.) the  red (blue) solid line shows the result of set A (B).
The dotted lines in both figures denote the results of the coupled-channel calculations  without the resonance
as corresponding to the solid lines in \figref{differ}.} \figlab{reson_differ}
\end{figure*}

\subsection{The $t$-dependence}

In this subsection, we discuss the $t$-dependence of the cross section. In
\figref{angle}
theoretical results for $d\sigma/dt (\theta = 0)$ with and without coupled channels
are compared with the LEPS
data~\cite{Mibe:2005er}. At forward angles, the differences between the tree level and the
coupled-channel results are not very large. In the backward region, however,
there are large differences. This is due to the fact that coupled-channel effects
tend to enhance the cross section at large angles. The effect is then further enlarged by the
inclusion of the $N^{*}$ resonance. Similar effects are also seen in
$\rho$-meson photoproduction, where coupled-channel effects successfully
reproduce the observed cross section at backward angles~\cite{Usov:2006wg}.
In our present calculation, the $N^{*}$ resonance has an important contribution such that the enhancement at
large angle is more prominent in the resonance energy region.

In the present calculation the forward angle structure is dominated by the Pomeron
exchange while at the backward angles the dominant contribution is caused by
coupled-channel effects with the nucleon resonance $N^{*}$. In many other hadronic reactions at energies
well above $2$~GeV one sees that the cross section at forward angles can be explained
quite accurately by Reggeon exchange~\cite{Guidal:1997hy, Corthals:2005ce, Mart:2004au, Toki:2007ab} while at backward angles more complicated
processes contribute.

\begin{figure*}
\begin{center}
\includegraphics[width=1.0 \textwidth]{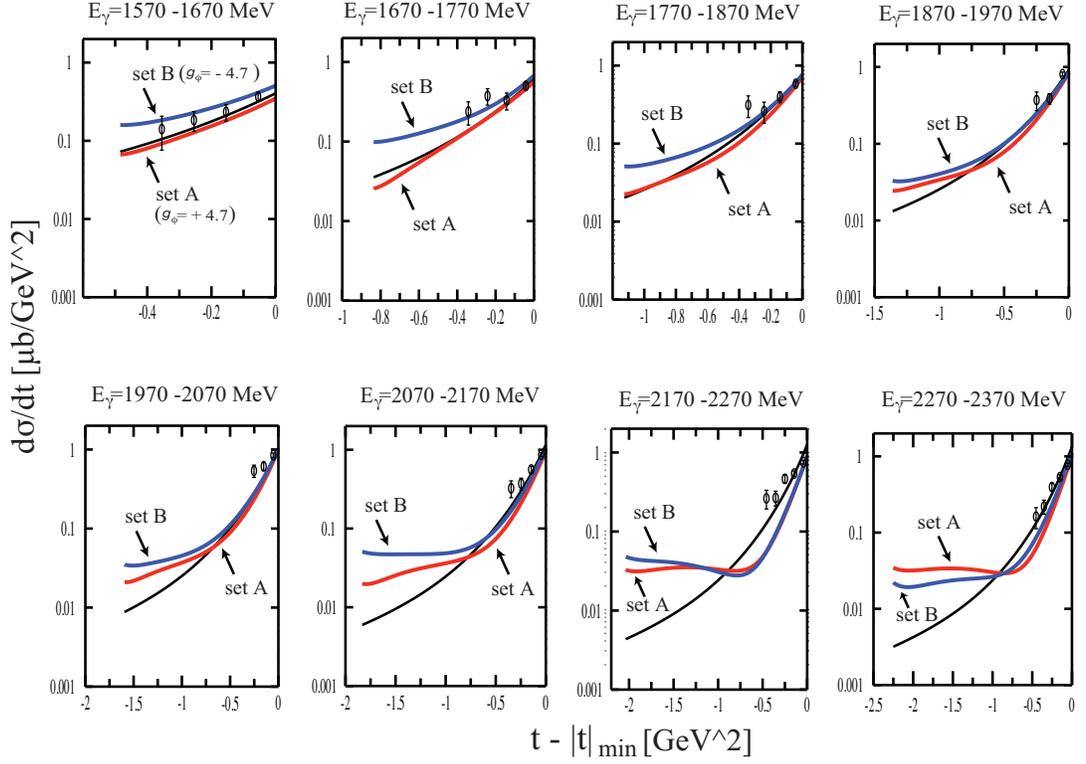}
\vskip -0.1in
\caption{[color online] The $t$-dependence of the cross section. The black
line is the tree level result and the red (blue) lines are results of set A (B).
The data are taken from Ref.~\cite{Mibe:2005er}.}
\figlab{angle}
\end{center}
\end{figure*}

\subsection{Spin-density matrix}

In general spin observables provide a sensitive test of the reaction mechanism.
We have therefore calculated these quantities to investigate the coupled-channel effects.
Following Ref.~\cite{Titov:1999eu, Schilling:1969um} we have calculated the spin-density matrices in the
Gottfried-Jackson (GJ) system.
In the LEPS experiment measurements have been done for $t+|t|_{min} > -0.2$ GeV$^{2}$.
In order to obtain the spin density matrices for a similar kinematics as for the LEPS experiment we have performed our calculation at
an angle of $\theta = 20$ degrees.

\begin{figure*}
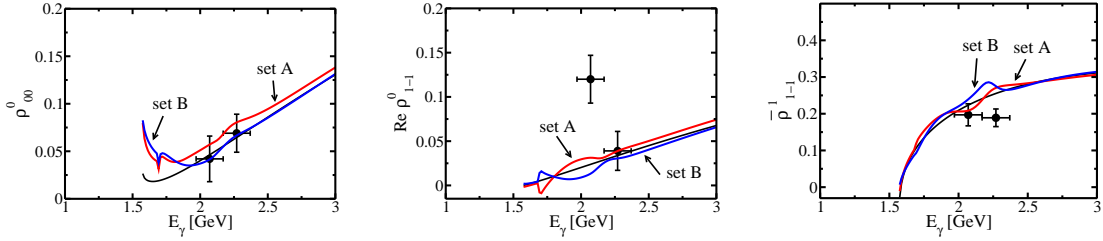

\begin{tabular}{ccc}
\begin{minipage}{0.3\hsize}
\begin{center}
\includegraphics[width=0.9 \textwidth]{rho_000.eps}
\end{center}
\end{minipage}
\begin{minipage}{0.3\hsize}
\begin{center}
\includegraphics[width=0.9 \textwidth]{rho_01-1.eps}
\end{center}
\end{minipage}
\begin{minipage}{0.3\hsize}
\begin{center}
\includegraphics[width=0.9 \textwidth]{rho_11-1.eps}
\end{center}
\end{minipage}
\end{tabular}
\caption{[color online] The spin-density matrix for $\phi$-meson
photoproduction as a function of the photon energy in the GJ system. In the calculation the $\phi$-meson angle is fixed at $\theta = 20$ degrees.
The black line is the tree level results and the red (blue) lines are results of set A (B). The data are taken from Ref.~\cite{Mibe:2005er}.}
\figlab{spin}
\end{figure*}

The l.h.s.~of \figref{spin} shows $\rho^{0}_{00}$ which is related to the
single spin-flip amplitude in the GJ system. At the tree level only the Pomeron
exchange can contribute to $\rho^{0}_{00}$ while the meson exchange
contributions vanish exactly.
The coupled-channel effects are large, as can be seen from the figure.
There is a wide structure in the energy region of $E_{\gamma} \simeq 1.8$ (set A) and $2$ GeV
(set B), and also a very narrow dip at the low energy $E_{\gamma} \sim 1.7$ GeV.
This dip structure is due to the contribution from the on-shell kaon in
the $K\Lambda(1520) \to \phi N$ channel.
Since the non spin-flip amplitude gives the dominant contribution to the cross
section at forward angles, the coupled-channel effects can be much more pronounced in
the spin-flip amplitude. The $N^{*}$ resonance gives only a minor contribution to
$\rho^{0}_{00}$, as shown by a tiny bump at $E_{\gamma} \sim 2.25$ GeV.

The middle panel of \figref{spin} shows $\rho^{0}_{1-1}$ which is related to the
double spin-flip amplitude in the GJ frame.
Only the Pomeron exchange contributes to this spin-density matrix at the tree level since meson exchanges
can not transfer spin two if they have spin less than two, and their
contributions thus vanish. The coupled-channel effects are very important especially for the $t$-channel
contribution of the $K\Lambda(1520) \to \phi N$ amplitude. We see either a dip (set A) or peak (set B)
structure at in low-energy region due to the on-shell kaon effect.
These effects are larger than the Pomeron contribution in the low-energy region.
The contribution due to the $N^{*}$ resonance is
small. Our calculation, however, does not reproduce the data as shown in the middle panel of
\figref{spin}.
From the figures of $\rho^{0}_{00}$ and $\rho^{0}_{1-1}$ we emphasize that higher order
perturbation effects are important for spin observables.
A similar conclusion is reached in Ref.~\cite{Ozaki:2007ka}.

The right panel of \figref{spin} shows $\rho^{1}_{1-1}$ which is related to the
asymmetry due to the interference of natural (Pomeron) and unnatural
($\pi$,$\eta$) parity $t$-channel exchanges. In this matrix element the non
spin-flip amplitude is dominant. The coupled-channel effect is small in $\rho^{1}_{1-1}$ due
to the dominance of the Pomeron contribution. There is, however, a visible contribution due
to the $N^{*}$ resonance.

\section{Conclusions}\seclab{concl}

We have investigated $\phi$-meson photoproduction motivated by the recent experimental observation of a peak/dip
structure near the threshold region. Our method is based on a coupled-channel K-matrix
approach. The kernel used in the K-matrix is constructed based on an effective
Lagrangian respecting the symmetries of QCD. In the present work we have focussed
on the role of the $K\Lambda(1520)$ reaction channel in coupled-channel
calculations because the threshold value of the channel is close to that of the
$\phi N$ channel. Additional interest in this channel comes from the fact that in
the $t$-channel contribution to the transition amplitude $
K\Lambda(1520) \to \phi N$, the exchanged kaon can be on-shell state in the energy region
where a distinct structure is observed in $\phi$-meson photoproduction.  The
coupled-channel effect driven by this $t$-channel contribution itself produces a peak
structure in the correct energy region.
The dominant contribution to the $\phi$-meson photoproduction
amplitude, however, is generated by the
Pomeron, compared to which the coupled-channel effect induced by the
$t$-channel in the $K\Lambda(1520) \to \phi N$ transition amplitude is insignificant.

As an alternative explanation for the observed structures we have considered the
effect of an $N^{*}$ resonance.
If the resonance contains an
hidden $s\bar{s}$ component (as would be the case for a $\phi N$ or a $K\Lambda(1520)$ molecular state)
it should have a large coupling to the
$K\Lambda(1520)$ and the $\phi N$ reaction channels.
For simplicity we
have considered only the case of a spin-$1/2$ resonance. We found that
destructive interference between the tree-level amplitude and the resonance coupled-channel contribution can produce
a dip structure which is in excellent agreement with the experimental data. These results suggest the existence of
an $N^{*}$, $J^{P} = 1/2^{-}$
resonance  with a mass of $2250$~MeV and width of about 100~MeV that has a large $s\bar{s}$ component.

To complete our calculations we have also presented our results for the
$t$-dependence of the $\phi$-meson photoproduction cross section and the
forward-angle spin-density matrices. In the cross section the coupled-channel effects are most
prominent at backward angles while they do contribute to certain polarization
observables at forward angles.
The $t$-dependence at large momentum transfer is perhaps one of clear cut signals of the $N^{*}$ resonance as well as the dip structure, and therefore it would be interesting if further comparison with experimental data will be available.

\appendix
\section{Effective Lagrangians and Form factors}\seclab{effL}
In the transition amplitude for $\gamma N \to \phi N$,
we calculate the pseudoscalar meson exchange amplitudes in terms of the following effective Lagrangians~\cite{Titov:2007fc}
\beq
\mathcal{L}_{\gamma \varphi \phi}
&=&\frac{eg_{\gamma \varphi \phi}}{m_{\varphi}}
\epsilon^{\mu \nu \alpha \beta}\partial_{\mu}\phi_{\nu}\partial_{\alpha}A_{\beta}\varphi, \\
\mathcal{L}_{\varphi NN}
&=&\frac{g_{\varphi NN}}{2M_{N}}\bar{N}\gamma_{\mu}\gamma_{5}N\partial^{\mu}\varphi
\eeq
where $\varphi$ stands for the pseudoscalar mesons ($\pi$, $\eta$).

We calculate the transition amplitudes for $\gamma N \to K\Lambda(1520)$ in terms of
the following effective Lagrangians~\cite{Nam:2005uq}
\beq
\mathcal{L}_{\gamma NN}
&=&
-e\bar{N}(\gamma^{\mu}-\frac{\kappa_{N}}{2M_{N}}\sigma^{\mu \nu}\partial_{\nu})NA_{\mu}, \\
\mathcal{L}_{\gamma KK}
&=&-ie(\partial^{\mu}K^{+}K^{-}-\partial^{\mu}K^{-}K^{+})A_{\mu}, \\
\mathcal{L}_{\gamma \Lambda^{*} \Lambda^{*}}
&=&-\frac{\kappa_{\Lambda^{*}}}{2M_{\Lambda^{*}}}
\bar{\Lambda}^{* \mu}\Slash{k}_{\gamma}\Slash{A}\Lambda^{*}_{\mu}, \\
\mathcal{L}_{KN\Lambda^{*}}
&=&\frac{g_{KN\Lambda^{*}}}{m_{K}}\bar{N}\gamma_{5}\partial_{\mu}K^{+}\Lambda^{* \mu}, \\
\mathcal{L}_{\gamma K \phi N}
&=& i\frac{eg_{KN\Lambda^{*}}}{m_{K}}\bar{N}\gamma_{5}K^{+}\Lambda^{* \mu}A_{\mu}.
\eeq
Coupling constants are shown in Table~\ref{couplings}.

\begin{table}[h]
\begin{tabular}{|c||c|c||c|}
\hline
 $g_{\gamma \pi \phi}$ & $-0.141$ & $g_{\gamma \eta \phi}$ & $-0.707$  \\
\hline
 $g_{\pi NN}$ & $13$ & $g_{\eta NN}$  & $1.94$ \\ \hline
$\kappa_{\Lambda^{*}}$ & $0.5$ & $g_{KN\Lambda^{*}}$ & $-11$ \\
\hline
\end{tabular}
\caption{Coupling constants}
\label{couplings}
\end{table}

In the amplitudes calculated from the above Lagrangians we have used a form factor,
\beq
F_{x}
&=& \frac{\Lambda_{c}^{4}}{(x-m_{x}^{2})^{2}+\Lambda_{c}^{4}} , \ x=s,t,u,
\eeq
where $\Lambda_{c}$ is the cut-off. We take the cut-off $\Lambda_{c} = 1.2$~GeV
for ground state hadrons and $\Lambda_{c} = 0.7$~GeV for the $\Lambda(1520)$ resonance. In
order to satisfy gauge invariance we follow the suggestion of Davidson and
Workman~\cite{Davidson:2001nn},
\begin{eqnarray*}
F\mathcal{M}
&=& F_{s}\mathcal{M}_{s,\circ}+F_{u}\mathcal{M}_{u,\circ}+F_{t}\mathcal{M}_{t,\circ} \\
&+& F_{c}(\mathcal{M}_{s,\times}+\mathcal{M}_{u,\times}+\mathcal{M}_{t,\times}+\mathcal{M}_{c}),
\end{eqnarray*}
where $\mathcal{M}_{x,\circ}$ is the gauge invariant part of the $x$-channel
amplitude and $\mathcal{M}_{x,\times}$ is not. Here $F_{c}$ is defined as
\beq
F_{c}
&=& 1-(1-F_{s})(1-F_{u})(1-F_{t}).
\eeq
We use this procedure for the $\gamma N \to K\Lambda(1520)$ kernel. For the $\phi
N \to K\Lambda(1520)$ kernel all born contributions are multiplied by the form
factor $F_{c}$. For the $\gamma N \to \phi N$ kernel we follow Titov's
approach~\cite{Titov:2007fc}.

\section{The determination of an effective kaon width}
In order to determine the effective width for the kaon exchanged in the transition $K\Lambda(1520) \to \phi N$,
we express the intermediate kaon propagator as a function of the CM energy $W$ and the angle $\theta_{K}$ between
 the out-going $\phi$-meson and in-coming kaon.
The left hand side of \eqref{ficwidth} can be described as
\begin{eqnarray*}
g_{\phi}(W, \rm{cos}\theta_{K})
&=& \frac{i}{(p_{\phi}-p_{K})^{2}-m_{K_{in}}^{2}+i\epsilon} \\
&=& \int dm_{\phi}^{2} \delta(m_{\phi}^{2}-\bar{m}_{\phi}^{2})
\frac{i}{(p_{\phi}-p_{K})^{2}-m_{K_{in}}^{2}+i\epsilon} \;,
\end{eqnarray*}
where $\bar{m}_{\phi}$ is the physical $\phi$-meson mass, $\bar{m}_{\phi} = 1.02$ GeV, and $m_{K_{in}}$ is the
exchanged kaon mass. By using a smearing function, we introduce the physical
$\phi$-meson width $\Gamma_{\phi}$
\begin{eqnarray*}
g_{\phi}(W, {\rm{cos}}\theta)
&\to& \int dm_{\phi}^{2} f(m_{\phi}^{2}-\bar{m}^{2}_{\phi}, \Gamma_{\phi})
\frac{i}{(p_{\phi}-p_{K})^{2}-m_{K_{in}}^{2}+i\epsilon} \;,
\end{eqnarray*}
where the smearing function $f$ is given by the imaginary part of the $\phi$-meson
propagator
\begin{eqnarray*}
f(m_{\phi}^{2}-\bar{m}_{\phi}^{2}, \Gamma_{\phi})
&=& \frac{\bar{m}_{\phi}\Gamma_{\phi}}
 {(m_{\phi}^{2}-\bar{m}_{\phi}^{2})^{2}+\bar{m}_{\phi}^{2}\Gamma_{\phi}^{2}}\;.
\end{eqnarray*}
When the exchanged kaon reaches the on-shell point, the propagator is given by
\begin{eqnarray*}
g_{\phi}(W, \rm{cos}\theta)
&\to& \int dm_{\phi}^{2} f(m_{\phi}^{2}-\bar{m}_{\phi}^{2},\Gamma_{\phi})
i(-i \delta(D({\rm{cos}}\theta, m_{\phi})) \;,
\end{eqnarray*}
where the function $D({\rm{cos}}\theta, m_{\phi})$ is defined by
\begin{eqnarray*}
D({\rm{cos}}\theta, m_{\phi})
&=& {\rm{cos}}\theta-F(m_{\phi}) \;.
\end{eqnarray*}
Here the function $F(m_{\phi})$ is given by
\beq
F(m_{\phi}) &=&
\frac{2E_{\phi}E_{K}-m_{\phi}^{2}}{2|\vec{p}_{\phi}||\vec{p}_{K}|} \;,
\eeq
where $\vec{p}_{\phi}$ and $\vec{p}_{K}$ are the 3-momentum vector of the
$\phi$-meson and the out-going kaon, respectively. We perform the integral,
giving
\begin{eqnarray*}
g_{\phi}(W, \rm{cos}\theta)
&=& \int \frac{dm_{\phi}^{2}}{dF}dF f(m_{\phi}-\bar{m}_{\phi},\Gamma_{\phi})
i(-i) \delta({\rm{cos}}\theta-F(m_{\phi})) \\
&=& \left( \frac{dF}{dm_{\phi}^{2}}\right)_{m_{\phi}^{2}=
 m^{'2}_{\phi}}^{-1} f(m_{\phi}^{'2}-\bar{m}_{\phi}^{2},\Gamma_{\phi}) \;.
\end{eqnarray*}
The introduced mass $m_{\phi}'$ denotes the solution of the equation
$D({\rm{cos}}\theta, m_{\phi}^{'})=0$ and is a function of ${\rm{cos}}\theta$. Along the
same lines one can also introduce the $\Lambda(1520)$ resonance width
$\Gamma_{\Lambda^{*}}$ in the picture,
\begin{eqnarray*}
g_{\Lambda^{*}}(W, \rm{cos}\theta)
&=& \int \frac{dm_{\Lambda^{*}}^{2}}{dF}dF f(m_{\Lambda^{*}}^{2}-\bar{m}_{\Lambda^{*}}^{2},\Gamma_{\Lambda^{*}})
i(-i) \delta({\rm{cos}}\theta-F(m_{\Lambda^{*}})) \\
&=& \left( \frac{dF}{dm_{\Lambda^{*}}^{2}}\right)_{m_{\Lambda^{*}}^{2}=
 m^{'2}_{\Lambda^{*}}}^{-1} f(m_{\Lambda^{*}}^{'2}-\bar{m}_{\Lambda^{*}}^{2},\Gamma_{\Lambda^{*}}) \;,
\end{eqnarray*}
where the mass $m_{\Lambda^{*}}'$ denotes the solution of the equation
${\rm{cos}}\theta-F(m_{\Lambda^{*}}^{'})=0$. Finally we introduce the effective kaon
width as
\begin{eqnarray*}
g_{K_{in}}(W, \rm{cos}\theta)
&=& \int \frac{dm_{K_{in}}^{2}}{dF}dF f(m_{K_{in}}^{2}-\bar{m}_{K_{in}}^{2},\Gamma_{K_{in}})
i(-i) \delta({\rm{cos}}\theta-F(m_{K_{in}})) \\
&=& \left( \frac{dF}{dm_{K_{in}}^{2}}\right)_{m_{K_{in}}^{2}=m^{'2}_{K_{in}}}^{-1} f(m_{K_{in}}^{'2}-\bar{m}_{K_{in}}^{2},\Gamma_{K_{in}}) \;,
\end{eqnarray*}
where the mass $m_{K_{in}}'$ refers to a solution of the equation
${\rm{cos}}\theta-F(m_{K_{in}})= 0$, and the function $F$ is defined as
\beq
F(m_{K_{in}}) &=&
\frac{2E_{\phi}E_{K}-m_{\phi}^{2}+(m_{K_{in}}^{2}-\bar{m}_{K}^{2})}{2|\vec{p}_{\phi}||\vec{p}_{K}|}
\;.
\eeq

We can obtain the ratio of the effective width of the intermediate kaon,
$K_{in}$, and the widths $\phi$-meson or the $\Lambda^*$-resonance from
$dF/dm_{x}$, where $x$ denotes $\phi, \Lambda^{*}$ or $K_{in}$,
\beq
\frac{(dF/dm_{\phi})|_{m_{\phi}=\bar{m}_{\phi}}}{(dF/dm_{K_{in}})|_{m_{K_{in}}=\bar{m}_{K_{in}}}} =
 \frac{dm_{K_{in}}}{dm_{\phi}} = \frac{\Gamma_{K_{in}}}{\Gamma_{\phi}}
\label{ratio_K_phi}
\eeq
and
\beq
\frac{(dF/dm_{\Lambda^{*}})|_{m_{\Lambda^{*}}=\bar{m}_{\Lambda^{*}}}}{(dF/dm_{K_{in}})|_{m_{K_{in}}=
 \bar{m}_{K_{in}}}} = \frac{dm_{K_{in}}}{dm_{\Lambda^{*}}} = \frac{\Gamma_{K_{in}}}{\Gamma_{\Lambda^{*}}} \;.
\label{ratio_K_Lst}
\eeq
\figref{width_ratio} shows $dm_{K_{in}}/dm_{\phi}$ and $dm_{K_{in}}/dm_{\Lambda^{*}}$
as functions of the photon energy $E_{\gamma}$.

\begin{figure*}
\begin{center}
\includegraphics[width=0.4 \textwidth]{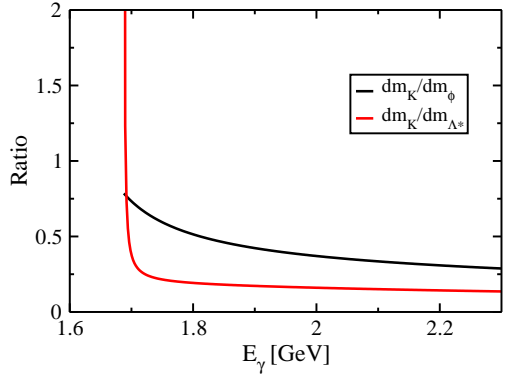}
\vskip -0.1in \caption{[color online] The ratio of the
widths given by Eqs.~(\ref{ratio_K_phi}),(\ref{ratio_K_Lst}). } \figlab{width_ratio}
\end{center}
\end{figure*}

Substituting the physical widths of the $\phi$-meson $\Gamma_{\phi} = 4.26$ MeV
and the $\Lambda(1520)$ resonance $\Gamma_{\Lambda^{*}} = 15.6$ MeV, we obtain from
\figref{width_ratio} the corresponding kaon width at $E_{\gamma} = 1.77$ GeV,
\begin{eqnarray*}
\Gamma_{K_{in}}^{\phi} = \frac{dm_{K_{in}}}{dm_{\phi}} \Gamma_{\phi}
=4.97 \ [\rm{MeV}] \\
\Gamma_{K_{in}}^{\Lambda^{*}} = \frac{dm_{K_{in}}}{dm_{\Lambda^{*}}} \Gamma_{\Lambda^{*}}
=9.87 \ [\rm{MeV}] \;,
\end{eqnarray*}
at $E_{\gamma} = 2.1$ GeV (the end of blue region) we obtain,
\begin{eqnarray*}
\Gamma_{K_{in}}^{\phi} = \frac{dm_{K_{in}}}{dm_{\phi}} \Gamma_{\phi}
=2.94 \ [\rm{MeV}], \\
\Gamma_{K_{in}}^{\Lambda^{*}} = \frac{dm_{K_{in}}}{dm_{\Lambda^{*}}} \Gamma_{\Lambda^{*}}
=7.23 \ [\rm{MeV}] \;.
\end{eqnarray*}

\begin{figure*}
\begin{tabular}{cc}
\begin{minipage}{0.5\hsize}
\begin{center}
\includegraphics[width=0.7 \textwidth]{g-function_for_phi_K_2050mev.eps}
\end{center}
\end{minipage}
\begin{minipage}{0.5\hsize}
\begin{center}
\includegraphics[width=0.7 \textwidth]{g_k_and_gLst_for_2050mev.eps}
\end{center}
\end{minipage}
\end{tabular}
\caption{[color online] The on-shell exchanged kaon propagator at $E_{\gamma} =
1.77$~GeV as a function of angle. In the upper figure, the black solid lines denote $g_{\phi}(\cos\theta)$
with the physical $\phi$-meson width $\Gamma_{\phi}$ and the red dashed lines are
$g_{K}(\cos\theta)$ with the corresponding kaon width $\Gamma^{\phi}_{K} = 4.97$~MeV.
In the lower figure, the black solid lines are $g_{\Lambda^{*}}(\cos\theta)$ with
the physical $\Lambda(1520)$ width $\Gamma_{\Lambda(1520)}$ and the red dashed
lines are $g_{K}(\cos\theta)$ with the corresponding kaon width
$\Gamma^{\Lambda^{*}}_{K} = 9.87$~MeV.} \figlab{g-funcions2.05}
\end{figure*}

\begin{figure*}
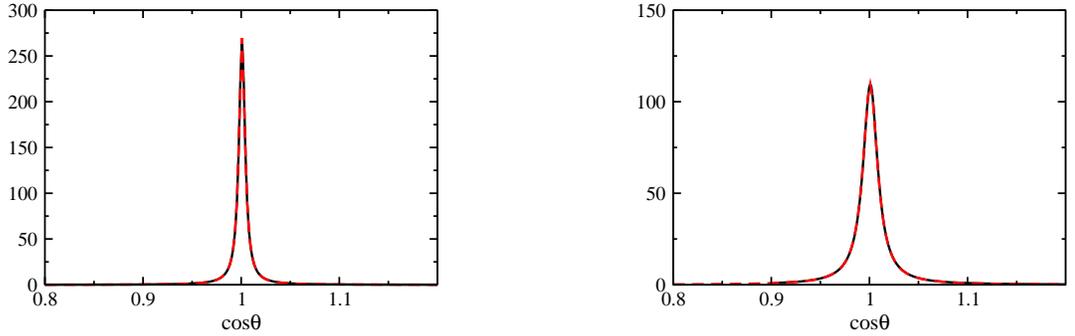

\begin{tabular}{cc}
\begin{minipage}{0.5\hsize}
\begin{center}
\includegraphics[width=0.7 \textwidth]{g-function2100_phi.eps}
\end{center}
\end{minipage}
\begin{minipage}{0.5\hsize}
\begin{center}
\includegraphics[width=0.7 \textwidth]{g-function2100_KLst.eps}
\end{center}
\end{minipage}
\end{tabular}
\caption{[color online] The on-shell exchanged kaon propagator at $E_{\gamma} =
2.1$~GeV as a function of angle. In the upper figure, the black solid lines are $g_{\phi}(\cos\theta)$ with
the physical phi width $\Gamma_{\phi}$ and the red dashed lines are
$g_{K}(\cos\theta)$ with the corresponding kaon width $\Gamma^{\phi}_{K} =
2.94$~MeV. In lower figure, the black solid lines are $g_{\Lambda^{*}}(\cos\theta)$ with
the physical $\Lambda(1520)$ width $\Gamma_{\Lambda(1520)}$ and the red dashed
lines are $g_{K}(\cos\theta)$ with the corresponding kaon width
$\Gamma^{\Lambda^{*}}_{K} = 7.23$~MeV.} \figlab{g-funcions2.2}
\end{figure*}

The functions $g$ are displayed at $E_{\gamma} = 1.77$~GeV in
\figref{g-funcions2.05} and for $E\gamma = 2.1$~GeV in \figref{g-funcions2.2}.
These figures show that $g_{K_{in}}(\cos\theta)$ is in good agreement with
$g_{\phi}(\cos\theta)$ and $g_{\Lambda^{*}}(\cos\theta)$. This means that the
effective width of the intermediate kaon can accurately account for the physical
decay width of the $\phi$-meson and the $\Lambda(1520)$-resonance. We define the
effective intermediate kaon width $\Gamma_{K}$ as
\beq
\Gamma_{K} &=& \sqrt{\Gamma_{K}^{\phi 2} + \Gamma_{K}^{\Lambda^{*} 2}} \;.
\eeq
Typical values for $\Gamma_{K}$ are for example
\begin{eqnarray*}
\Gamma_{K}
&=& 11.1 \ [\rm{MeV}] \ at \ E_{\gamma} = 1.77 \ [\rm{GeV}] \;, \\
\Gamma_{K}
&=& 7.79 \ [\rm{MeV}] \ at \ E_{\gamma} = 2.1 \ [\rm{GeV}] \;.
\end{eqnarray*}

\section{Rarita-Schwinger vector spinor}
We can decompose the Rarita-Schwinger vector spinor~\cite{Rarita:1941mf} for the different spin states
as follows,
\begin{eqnarray*}
u^{\mu}(p,3/2)
&=& e^{\mu}_{+}(p)u(p,1/2), \\
u^{\mu}(p,1/2)
&=& \sqrt{\frac{2}{3}}e^{\mu}_{0}(p)u(p,1/2)+\sqrt{\frac{1}{3}}e^{\mu}_{+}(p)u(p,-1/2), \\
u^{\mu}(p,-1/2)
&=& \sqrt{\frac{1}{3}}e^{\mu}_{-}(p)u(p,1/2)+\sqrt{\frac{2}{3}}e^{\mu}_{0}(p)u(p,-1/2), \\
u^{\mu}(p,-3/2)
&=& e^{\mu}_{-}(p)u(p,-1/2).
\end{eqnarray*}
Here we use the basis four-vectors $e^{\mu}_{\lambda}$ which are given as
\begin{eqnarray*}
e^{\mu}_{\lambda}(p)
&=& \left( \frac{\vec{\epsilon}_{\lambda}\cdot \vec{p}}{M_{B}}, \ \epsilon_{\lambda}+\frac{\vec{p}(\vec{\epsilon}_{\lambda}\cdot \vec{p})}{M_{B}(p^{0}+M_{B})} \right) \ {\rm{with}}
\end{eqnarray*}
\begin{eqnarray*}
\vec{\epsilon}_{+}=-\frac{1}{\sqrt{2}}(1,i,0), \ \vec{\epsilon}_{0}=(0,0,1), \ \vec{\epsilon}_{-}=\frac{1}{\sqrt{2}}(1,-i,0).
\end{eqnarray*}

\section{SPIN DENSITY MATRIX}
In terms of helicity amplitudes $T_{\lambda_{f},\lambda; \lambda_{i}, \lambda_{\gamma}}$,
spin density matrices can be written as \cite{Titov:1999eu, Schilling:1969um}
\beq
\rho^{0}_{\lambda \lambda^{'}}
&=& \frac{1}{N}\sum_{\lambda_{\gamma}, \lambda_{i}, \lambda_{f}}
T_{\lambda_{f},\lambda; \lambda_{i}, \lambda_{\gamma}} T^{*}_{\lambda_{f},\lambda^{'}; \lambda_{i}, \lambda_{\gamma}}, \\
\rho^{1}_{\lambda \lambda^{'}}
&=& \frac{1}{N}\sum_{\lambda_{\gamma}, \lambda_{i}, \lambda_{f}}
T_{\lambda_{f},\lambda; \lambda_{i}, -\lambda_{\gamma}} T^{*}_{\lambda_{f},\lambda^{'}; \lambda_{i}, \lambda_{\gamma}}, \\
\rho^{2}_{\lambda \lambda^{'}}
&=& \frac{i}{N}\sum_{\lambda_{\gamma}, \lambda_{i}, \lambda_{f}}\lambda_{\gamma}
T_{\lambda_{f},\lambda; \lambda_{i}, -\lambda_{\gamma}} T^{*}_{\lambda_{f},\lambda^{'}; \lambda_{i}, \lambda_{\gamma}}, \\
\rho^{3}_{\lambda \lambda^{'}}
&=& \frac{1}{N}\sum_{\lambda_{\gamma}, \lambda_{i}, \lambda_{f}}\lambda_{\gamma}
T_{\lambda_{f},\lambda; \lambda_{i}, \lambda_{\gamma}} T^{*}_{\lambda_{f},\lambda^{'}; \lambda_{i}, \lambda_{\gamma}},
\eeq
where
\beq
N
&=& \sum |T_{\lambda_{f},\lambda; \lambda_{i}, \lambda_{\gamma}}|^{2}.
\eeq
Here the helicities $\lambda_{\gamma}$, $\lambda_{i}$ and $\lambda_{f}$ are for the photon, initial and final nucleons,
while $\lambda$ and $\lambda^{'}$ are for the $\phi$-meson.
In order to  calculate the spin density matrix in the GJ system, one need a transformation from the CM to the GJ system.
This transformation is done in terms of
\beq
T^{GJ}_{\lambda_{f}, \lambda{\phi}; \lambda_{i}, \lambda_{\gamma}}
&=& \sum_{l, m} T^{CM}_{\lambda_{f} l ; m \lambda_{\gamma}} d^{1}_{l \lambda}(-\omega_{\phi})d^{1/2}_{m \lambda_{i}}(-\omega_{p}),
\eeq
where the corresponding Wigner rotating angles are given by
\beq
\omega_{\phi}
&=& {\rm{a cos}}\left( \frac{{\rm{cos}} \theta - u_{\phi}}{1 - u_{\phi}{\rm{cos}} \theta} \right), \\
\omega_{p}
&=& {\rm{a tan}} \left( \frac{u_{\phi} {\rm{sin}}(\pi-\theta) (1-v^{2}_{p})^{1/2}}{v_{p}+u_{\phi}{\rm{cos}}(\pi-\theta)}\right).
\eeq
Here $\theta$ is the angle of the out going $\phi$-meson in the c.m.s.\ while $v_{p}$ and $u_{\phi}$ are the proton and
the $\phi$-meson velocities in the c.m.s.\ respectively.



\begin{thebibliography}{99}


\bibitem{Nakano:2003qx}
  T.~Nakano {\it et al.}  [LEPS Collaboration],
  Phys.\ Rev.\ Lett.\  {\bf 91}, 012002 (2003)
  [arXiv:hep-ex/0301020].

\bibitem{Nakano:2008ee}
  T.~Nakano {\it et al.}  [LEPS Collaboration],
  Phys.\ Rev.\  C {\bf 79}, 025210 (2009)
  [arXiv:0812.1035 [nucl-ex]].

\bibitem{Barrow:2001ds}
  S.~P.~Barrow {\it et al.}  [Clas Collaboration],
  Phys.\ Rev.\  C {\bf 64}, 044601 (2001)
  [arXiv:hep-ex/0105029].

\bibitem{Niiyama:2008rt}
  M.~Niiyama {\it et al.},
  Phys.\ Rev.\  C {\bf 78}, 035202 (2008)
  [arXiv:0805.4051 [hep-ex]].

\bibitem{Anciant:2000az}
  E.~Anciant {\it et al.}  [CLAS Collaboration],
  Phys.\ Rev.\ Lett.\  {\bf 85}, 4682 (2000)
  [arXiv:hep-ex/0006022].

\bibitem{Mibe:2005er}
  T.~Mibe {\it et al.}  [LEPS Collaboration],
  Phys.\ Rev.\ Lett.\  {\bf 95}, 182001 (2005)
  [arXiv:nucl-ex/0506015].

\bibitem{Santoro:2008ai}
  J.~P.~Santoro {\it et al.}  [CLAS Collaboration],
  Phys.\ Rev.\  C {\bf 78}, 025210 (2008)
  [arXiv:0803.3537 [nucl-ex]].

\bibitem{Durham}
Durham data base. http://www.slac.stanford.edu/spires/hepdata/.

\bibitem{Titov:2003bk}
  A.~I.~Titov and T.~S.~H.~Lee,
  Phys.\ Rev.\  C {\bf 67}, 065205 (2003)
  [arXiv:nucl-th/0305002].

\bibitem{Titov:2007fc}
  A.~I.~Titov and B.~Kampfer,
  Phys.\ Rev.\  C {\bf 76}, 035202 (2007)
  [arXiv:0705.2010 [nucl-th]].

 \bibitem{Titov:1999eu}
  A.~I.~Titov, T.~S.~Lee, H.~Toki and O.~Streltsova,
  Phys.\ Rev.\  C {\bf 60}, 035205 (1999).

  \bibitem{Nam:2005uq}
  S.~I.~Nam, A.~Hosaka and H.~C.~Kim,
  Phys.\ Rev.\  D{\bf 71}, 114012 (2005)
  [arXiv:hep-ph/0503149].

\bibitem{Korchin:1998ff}
  A.~Y.~Korchin, O.~Scholten and R.~G.~E.~Timmermans,
  Phys.\ Lett.\  B {\bf 438}, 1 (1998)
  [arXiv:nucl-th/9811042].

\bibitem{Usov:2005wy}
  A.~Usov and O.~Scholten,
  Phys.\ Rev.\  C {\bf 72}, 025205 (2005)
  [arXiv:nucl-th/0503013].

\bibitem{Scholten:2002tn}
  O.~Scholten, S.~Kondratyuk, L.~Van Daele, D.~van Neck, M.~Waroquier and A.~Y.~Korchin,
  Acta Phys.\ Polon.\  B {\bf 33}, 847 (2002).

\bibitem{Donnachie:1987pu}
  A.~Donnachie and P.~V.~Landshoff,
  Phys.\ Lett.\  B {\bf 185}, 403 (1987).

\bibitem{Muramatsu:2009zp}
  N.~Muramatsu {\it et al.},
  arXiv:0904.2034 [nucl-ex].

\bibitem{Barber:1980zv}
  D.~P.~Barber {\it et al.},
  Z.\ Phys.\  C {\bf 7}, 17 (1980).

\bibitem{KanadaEn'yo:2008wm}
  Y.~Kanada-En'yo and D.~Jido,
  Phys.\ Rev.\  C {\bf 78}, 025212 (2008)
  [arXiv:0804.3124 [nucl-th]].

\bibitem{Jido:2008kp}
  D.~Jido and Y.~Kanada-En'yo,
  Phys.\ Rev.\  C {\bf 78}, 035203 (2008)
  [arXiv:0806.3601 [nucl-th]].


\bibitem{Usov:2006wg}
  A.~Usov and O.~Scholten,
  Phys.\ Rev.\  C {\bf 74}, 015205 (2006)
  [arXiv:nucl-th/0604009].

\bibitem{Guidal:1997hy}
  M.~Guidal, J.~M.~Laget and M.~Vanderhaeghen,
  Nucl.\ Phys.\  A {\bf 627}, 645 (1997).

\bibitem{Corthals:2005ce}
  T.~Corthals, J.~Ryckebusch and T.~Van Cauteren,
  Phys.\ Rev.\  C {\bf 73}, 045207 (2006)
  [arXiv:nucl-th/0510056].

\bibitem{Mart:2004au}
  T.~Mart and C.~Bennhold,
  arXiv:nucl-th/0412097.

\bibitem{Toki:2007ab}
  H.~Toki, C.~Garcia-Recio and J.~Nieves,
  Phys.\ Rev.\  D {\bf 77}, 034001 (2008)
  [arXiv:0711.3536 [hep-ph]].

\bibitem{Schilling:1969um}
  K.~Schilling, P.~Seyboth and G.~E.~Wolf,
  Nucl.\ Phys.\  B {\bf 15}, 397 (1970)
  [Erratum-ibid.\  B {\bf 18}, 332 (1970)].

\bibitem{Ozaki:2007ka}
  S.~Ozaki, H.~Nagahiro and A.~Hosaka,
  Phys.\ Lett.\  B {\bf 665}, 178 (2008)
  [arXiv:0710.5581 [hep-ph]].

\bibitem{Rarita:1941mf}
  W.~Rarita and J.~Schwinger,
  Phys.\ Rev.\  {\bf 60}, 61 (1941).

\bibitem{Davidson:2001nn}
  R.~M.~Davidson and R.~Workman,
  Phys.\ Rev.\  C {\bf 63}, 058201 (2001)
  [arXiv:nucl-th/0102046].


\end{thebibliography}
\end{document}